\documentclass[a4paper,fleqn,usenatbib,useAMS,onecolumn]{mnras}
\usepackage{geometry}                
\usepackage{amssymb}
\usepackage{amsmath}
\usepackage{graphicx}
\usepackage{epsfig}
\usepackage{epstopdf}

\newcommand{\bs}{{\boldsymbol\xi}}
\newcommand{\bu}{{\bf u}}
\newcommand{\op}{{{\mathcal{L}}}}
\newcommand{\curlF}{{{\mathcal{F}}}}
\newcommand{\measur}{{{\Lambda}}}
\newcommand{\kerne}{{{\mathcal{K}}}}
\newcommand{\nmode}{\omega_{\rm ref}}
\newcommand{\dU}{{\dot U}}

\newcommand{\dV}{{\dot V}}
\newcommand{\pth}{{\partial_\theta}}
\newcommand{\br}{{\bf r}}
\newcommand{\cosec}{{\rm cosec\,}}
\newcommand{\bT}{{\bf T}}
\newcommand{\bnabla}{{\boldsymbol\nabla}}
\newcommand{\rhat}{{\bf\hat r}}

\newcommand{\that}{{\boldsymbol{\hat\theta}}}
\newcommand{\phat}{{\boldsymbol{\hat\phi}}}
\title[Sensitivity of normal-mode coupling measurements]{Sensitivity of Helioseismic Measurements of Normal-mode Coupling to Flows and Sound-speed Perturbations}
\author[S. M. Hanasoge et al.]{Shravan M. Hanasoge$^{1,6}$\thanks{Contact e-mail: \href{mailto:hanasoge@tifr.res.in}{hanasoge@tifr.res.in}}, Martin Woodard$^2$, H. M. Antia$^1$,\newauthor Laurent Gizon$^{3,4,6}$, \& Katepalli R. Sreenivasan$^{5,6}$
\\
$^{1}${Department of Astronomy \& Astrophysics, Tata Institute of Fundamental Research, Mumbai, India}\\
$^{2}${North-West Research Associates, Inc., Boulder, Colorado}\\
$^{3}${Max-Planck Institut f\"{u}r Sonnensystemforschung, G\"{o}ttingen, Germany}\\
$^{4}${Department of Astronomy, University of G\"{o}ttingen, G\"{o}ttingen, Germany}\\
$^{5}${New York University, New York, USA}\\
$^{6}${Centre for Space Science, New York University, Abu Dhabi, UAE}
}
\date{}

\pubyear{2017}

\begin{document}
\label{firstpage}
\pagerange{\pageref{firstpage}--\pageref{lastpage}}
\maketitle


\begin{abstract}In this article, we derive and compute the sensitivity of measurements of coupling between normal modes of oscillation in the Sun to underlying flows. The theory is based on first-Born perturbation theory, and the analysis is carried out using the formalism described by \citet{lavely92}. Albeit tedious, we detail the derivation and compute the sensitivity of specific pairs of coupled normal modes to anomalies in the interior. Indeed, these kernels are critical for the accurate inference of convective flow amplitudes and large-scale circulations in the solar interior. We resolve some inconsistencies in the derivation of \citet{lavely92} and reformulate the fluid-continuity condition. We also derive and compute sound-speed kernels, paving the way for inverting for thermal anomalies alongside flows. 
\end{abstract}

\begingroup
\let\clearpage\relax
\endgroup
\newpage
\maketitle

\section{Introduction}
The use of wavefield correlations to infer internal structure has a long history in seismology \citep[e.g.][]{dahlen68, woodhouse80, lavely92, DT98,woodard06,woodard14, woodard16}. The normal modes of oscillation are typically calculated with respect to a given model of the Sun (generally non-magnetic, non-rotating, spherically symmetric but not necessarily so), and in this state the modes are deemed ``uncoupled". However, given that our knowledge of the Sun's interior is incomplete, there will be deviations in our model from reality, causing the modes to become coupled. For a temporally static perturbation to a given model, scattering can only occur across wavenumber, redistributing modal power in a manner that depends on the type of perturbation. The redistributed power is quantified by correlating the modes (i.e. spherical-harmonic time-series filtered around the corresponding resonant frequency) and relating the coupling to internal structure.

The study of convective flows in the solar interior is a topic of major interest in contemporary helioseismology. Recent measurements using time-distance helioseismogy \citep{duvall} have suggested that convective-flow amplitudes are significantly weaker than theory and numerical simulations suggest \citep{hanasoge12_conv}. However, in contradiction to these results, \citet{greer15} measured large convective velocities using ring-diagram analysis \citep{hill88}, in agreement with convection simulations \citep[ASH,][]{miesch_etal_08}. In particular, \citet{woodard16} used mode-coupling theory to interpret correlations of Michelson Doppler Imager data \citep[MDI;][]{scherrer95} of spherical harmonic time series across different azimuthal orders $m$ at the same harmonic degrees $\ell$. The conclusion of the \citet{woodard16} study was that the amplitudes are larger by an order in magnitude than those \citet{hanasoge12_conv} obtained. In addition to understanding the physics of solar thermal transport, the resolution of this issue of convective-flow amplitudes is a critical to progress in related areas such as angular momentum transport and the solar dynamo \citep[e.g.][]{arfm2016}.

The simplicity of the data-handling procedure and elegance of the underlying interpretive theory make mode-coupling analysis a very attractive technique to study this flow inference problem. In the \citet{woodard16} study, the inference of flow velocity magnitude was empirical and inversions to obtain flow magnitudes were not carried out. 
To that end, this manuscript aims to build the theory for performing inversions using these correlations. 

Although set out in \citet{lavely92}, typographical errors appear in the expressions for kernels, specifically the lack of consistency between the toroidal-flow kernel in their equations (D20) and (C34) and to a lesser extent, a missing factor $i$ in their equation (C27). Moreover,  the condition on flow conservation should limit the number of free scalar functions to two (as opposed to three), but \citet{lavely92} write the inverse problem in terms of three scalar functions. To test the expressions for all kernels and address the issue of flow conservation, we set out to derive them here in the context of normal-mode measurements that are currently available from the Helioseismic and Magnetic Imager \citep[HMI;][]{hmi} and MDI missions.

\section{Definitions and helioseismic measurements}\label{defsand}
The Lagrangian-displacement eigenfunction $\bs_k$ of a non-rotating, non-attenuating, non-magnetic, spherically symmetric solar model, obeys \citep[e.g.][]{jcd_notes}
\begin{equation}
-\rho\omega_k^2 \bs_k -\bnabla(\rho c^2 \bnabla\cdot\bs_k - \rho g\, \bs_k\cdot\rhat) - g\rhat\bnabla\cdot(\rho\bs_k) =0,\label{fullop}
\end{equation}
where $\omega_k$ is the real resonant frequency, $\rho$ the density, $c$ the sound speed, $\rhat$ the radially outward unit vector and $g$ the gravity. In the following analysis, we assume that the density is only a function of radius and is entirely determined (therefore not a function of horizontal coordinate or time).{ We have invoked Cowling's approximation \citep[e.g.][]{jcd_notes} whereby perturbations to the gravitational potential of the Sun induced by waves are ignored. This is justified because we are primarily interested in waves that propagate in the convection zone which are well modelled in the Cowling limit (only very low-degree modes that traverse the radiative interior and core significantly perturb the gravitational potential). However, the present treatment may also be extended to the full oscillation equations, e.g. \citet{lavely92}.}
The operator~(\ref{fullop}) is Hermitian \citep[e.g.][]{ostriker67} and therefore the eigenfrequencies and eigenfunctions for that operator are real. 
Note that for the sake of notational convenience, $k = (\ell,m,n)$ is used to denote a specific mode, typically categorised by three quantum numbers, spherical harmonic degree $\ell$, azimuthal order $m$ and radial order $n$. Although for non-rotating, spherically symmetric models, the modes are degenerate in $m$, i.e. $\omega_{\ell, m,n} = \omega_{\ell,0,n}$, we retain the dependence on $m$ for the sake of completeness here. 
We rewrite equation~(\ref{fullop}) for the sake of brevity as  
\begin{equation}
(-\rho\,\omega_k^2 + \op_0)\bs_k = 0. \label{shorthand}
\end{equation}
The eigenfunctions $\bs_k$ satisfy the following orthonormality relationship
\begin{equation}
\int_\odot d\br\,\rho\,\bs^*_k\cdot\bs_{k'} = \delta_{kk'},
\end{equation}
{ where $d\br = r^2\,\sin\theta\,dr\,d\theta\,d\phi$ is an infinitesimal volume element, $r$ is radius, $\theta$ is co-latitude and $\phi$ is longitude.} This is an important property for the analysis to follow. 

Because the eigenfunctions are a complete set \citep[with zero Lagrangian-pressure perturbation on the outer boundary, e.g.][]{jcd_notes}, a general wavefield $\bxi(\br,\omega)$ may be written as 
\begin{equation}\bxi(\br,\omega) = \sum_q a_q(\omega)\,\bs_q(\br) = \sum_q a^\omega_q\,\bs_q(\br),\label{genwave}\end{equation} where { $\omega$ is temporal frequency,} $a_q$ are coefficients representing the phase and amplitude of the $q$th mode and we write $a^\omega_q =  a_q(\omega)$ for the sake of notational convenience. { The governing equation for driven, undamped oscillations is given by 
\begin{equation}
(-\rho\omega^2 + \op_0)\bxi = {\bf J}, \label{eq0}
\end{equation}
where ${\bf J}$ is the unperturbed source, $\bxi$ the zeroth-order wavefield and $\op_0$ is the unperturbed operator  \citep[whose properties are derived from a basic solar model such as model S;][]{jcd} as defined in equation~(\ref{shorthand}).} Note that the MDI and HMI missions provide time series of spherical-harmonic coefficients of oscillations, i.e. $a_k(\omega)$, measured at the solar photosphere. { In practice, because we only observe the side of the Sun that faces us, it is not possible to precisely spatially decompose the measured oscillation signals into individual spherical harmonics. {  This partial visibility leads to {\it leakage} between modes across $\ell$, introducing systematical errors in eventual inferences \citep[e.g.][]{larson15, woodard16}. Further, at a given frequency (between ridges for instance), many different radial orders contribute, leading to diminished ability in distinguishing between them. Temporal gaps in observations, due to instrumental or other issues, lead to mixing across temporal frequencies (and hence radial orders), thereby exacerbating this issue. To mitigate the problem of radial-order leakage, we use measurements at frequencies within a linewidth of resonance, ensuring that the associated radial order contributes a dominant fraction of observed mode power. Nevertheless, leakage across radial orders and spherical-harmonic degrees will systematically influence the results of mode-coupling analyses and need to be carefully considered.}}

\subsection{Perturbation theory} { The effects of rotation, flows, magnetic fields or thermal asphericities are modelled by adding a perturbation to the operator $\delta\op$ to $\op_0$ in equation~(\ref{eq0}), in turn leading to a perturbation to the wavefield, $\delta\bxi$. Ignoring perturbations to the source ${\bf J}$, we obtain  
\begin{eqnarray}
(-\rho\,\omega^2  + \op_0 +\delta\op)(\bxi + \delta\bxi) = {\bf J},\label{eq.pert}
\end{eqnarray}
for the wavefield in a perturbed Sun. For time-varying perturbations,} $\delta\op$ behaves as a convolution operator. To first order, we have therefore
\begin{eqnarray}
(-\rho\,\omega^2  + \op_0)\delta\bxi = - \int d\omega'\,\delta\op_{\omega-\omega'}\bxi(\br,\omega'),\label{eq.pert2}
\end{eqnarray}
where the convolution over frequency captures the time variability of the perturbation operator $\delta\op$ (Eq.~[\ref{convint}] of Appendix~\ref{fourier}).  
Perturbing equation~(\ref{genwave}) while assuming that the eigenfunctions are unmodified by the perturbation, we may express the change to the wavefield as a linear combination of the original set of eigenfunctions,
\begin{equation}
\delta\bxi = \sum_{j} \delta a^\omega_j\,\bs_j. 
\end{equation}
Using equation~(\ref{shorthand}), the left hand side of equation~(\ref{eq.pert}) may be simplified thus
\begin{equation}
(-\rho\omega^2 + \op_0)\delta\bxi = \rho\,\sum_j  (\omega_j^2 - \omega^2)\,\delta a^\omega_j\, \bs_j.\label{pertexp}
\end{equation} 
Solar modes are subject to a small amount of attenuation $\gamma_k\ll\omega_k$ that perturb the eigenfunctions and eigenfrequencies (obtained from Eq.~[\ref{fullop}]) by introducing small imaginary components. However, because the attenuation term is generally much smaller than the resonant frequency, we only consider it as a perturbation to the resonant frequency and ignore changes to the eigenfunctions. { The inclusion of damping is driven to a large extent by the desire to apply the theoretical model to the analysis of oscillation data. Therefore, we substitute $\omega_k$ by $\bar\omega_k = \omega_k- i\gamma_k/2$ and continue to treat $\bs_k$ as a real quantity.} 
The sign of the attenuation term is negative owing to the Fourier convention we adopt (Appendix~\ref{fourier}).
Dotting both sides of equation~(\ref{eq.pert2}) by $\bs^*_{k}$, using equation~(\ref{pertexp}) and integrating over the solar volume, we obtain
\begin{equation}
\sum_j \delta a^\omega_j\, (\bar\omega_j^2 - \omega^2)\int_\odot d\br\, \rho\,\bs^*_{k}\cdot\bs_j =  -\sum_{k'}\int_\odot d\br\,\int d\omega' \,a^{\omega'}_{k'} \,\bs^*_{k}\cdot\delta\op_{\omega-\omega'}\, \bs_{k'}.
\end{equation}
This gives us the coupling between modes $k$ and all of $k'$,
\begin{equation}
\delta a^\omega_{k} =  -\frac{1}{(\bar\omega_{k}^2 - \omega^2)}\sum_{k'}\int d\omega' a^{\omega'}_{k'} \,\int_\odot d\br\, \bs^*_{k}\cdot\delta\op_{\omega - \omega'}\bs_{k'}.\label{coupling}
\end{equation}
We quantify the spectral profile of the mode thus,
\begin{equation}
R^{\omega}_{k} = \frac{1}{\bar\omega_{k}^2 - \omega^2}.
\label{couplingcon}
\end{equation}
The term in the denominator of equation~(\ref{couplingcon}) suggests that the coupling is strongest when the temporal frequency $\omega$ is very close to the resonant mode frequency and falls rapidly as the frequency difference grows. { The modification of the frequency through the introduction of a small imaginary component (as described above) removes the singularity associated with the point $\omega = \omega_k$ in equation~(\ref{couplingcon}).}
The coupling integral between the two eigenstates $k$ and $k'$ is defined thus,
\begin{equation}
\Lambda^{k}_{k'}(\sigma) = -\int_\odot d\br \,\,\bs^*_{k}\cdot\delta\op_{\sigma}\,\bs_{k'}.\label{coupledef}
\end{equation}
Equation~(\ref{coupling}) may thus be written as
\begin{equation}
\delta a^{\omega}_{k} = R^{\omega}_{k} \sum_{k'} \int d\omega' \,a^{\omega'}_{k'} \Lambda^{k}_{k'}(\omega-\omega') = R^{\omega}_{k} \sum_{k'} \int d\omega' \,a^{\omega'}_{k'} \Lambda^{k'*}_k(\omega'-\omega),\label{coupling.developed}
\end{equation}
where for the perturbation operators of interest here, i.e. for sound-speed and flows, it is shown in Appendix~\ref{adjoint.props} that $\Lambda^{k'*}_k(\omega'-\omega) = \Lambda^{k}_{k'}(\omega-\omega')$. 
 We therefore consider the complex conjugate of equation~(\ref{coupling.developed})
\begin{equation}
\delta a^{\omega^*}_{k} = R^{\omega*}_{k} \sum_{k'} \int d\omega' \,a^{\omega'*}_{k'}\,\Lambda^{k'}_k(\omega'-\omega).\label{firstcut} 
\end{equation}
%
We now multiply both sides of equation~(\ref{firstcut}) by $a^{{\omega+\sigma}}_q$, ensemble average and note that the excitation of waves in the Sun is well approximated as uncorrelated across modes \citep[e.g.][]{woodard07}. We have $\langle a^{\omega + \sigma}_q\, a^{\omega'*}_{k}\rangle = N_q\,|R^{\omega'}_q|^2 \delta(\omega + \sigma - \omega')\, \delta_{qk}$, where the angular brackets indicate ensemble averaging and $N_j$ is the mode-amplitude normalization. 
Equation~(\ref{firstcut}) is simplified thus
\begin{equation}
\langle a^{{\omega+\sigma}}_q\,\delta a^{\omega*}_{k} \rangle= R^{\omega*}_{k} \sum_{k'}  \int d\omega'\, \langle a^{\omega+\sigma}_q\,a^{\omega'*}_{k'}\rangle\,\Lambda^{k'}_k(\omega'-\omega) 
=N_q\,R^{\omega*}_{k}\, |R^{\omega+\sigma}_{q}|^2\, \Lambda^{q}_k(\sigma).\label{pert1}
\end{equation}
Similarly,
\begin{equation}
\langle \delta a^{{\omega+\sigma}}_q\, a^{\omega*}_{k} \rangle = N_{k}\,R^{(\omega+\sigma)}_{q}\, |R^{\omega}_{k}|^2\, \Lambda^{k*}_{q}(-\sigma),\label{pert2}
\end{equation}
and using equations~(\ref{pert1}) and~(\ref{pert2}), we model the expectation value of the cross-spectral correlation signal,
\begin{equation}
\langle a^{{\omega+\sigma}}_q\,\delta a^{\omega*}_{k} + \delta a^{{\omega+\sigma}}_q\, a^{\omega*}_{k} \rangle= (N_q\,R^{\omega*}_{k}\, |R^{\omega+\sigma}_{q}|^2 + N_{k}\,R^{(\omega+\sigma)}_{q}\, |R^{\omega}_{k}|^2) \Lambda^{q}_k(\sigma) = H\,\Lambda^{q}_k(\sigma),\label{connect}
\end{equation}
where
\begin{equation}
H(k,q,\sigma) = N_q\,R^{\omega*}_{k}\, |R^{\omega+\sigma}_{q}|^2 + N_{k}\,R^{(\omega+\sigma)}_{q}\, |R^{\omega}_{k}|^2.\label{defineh}
\end{equation}
{ Equations~(\ref{connect}) and~(\ref{defineh}) define the essence of normal-mode-coupling problem in helioseismology. In the remaining part of this article, we focus on calculating the coupling term $\Lambda^q_k$ and expressing it in terms of the underlying perturbations.}

\subsection{Basis expansion of perturbations and eigenfunctions}
Making use of terminology originally set out in \citet{lavely92}, we first introduce spherical harmonics $Y_\ell^m(\theta, \phi)$, the basis set on which eigenfunctions are projected. The generalised spherical harmonic $Y_\ell^{Nm} = D^\ell_{Nm}(\phi,\theta,0)$, where $D^\ell_{Nm}$ is the Wigner rotation matrix that relates spherical harmonics in rotated frames, will also play an important role in the analysis.

The mode eigenfunction describing oscillations in a non-rotating, non-magnetized, spherically symmetric model of the Sun may be written using spheroidal harmonics thus $\bs_k = U(r)\,Y^m_\ell\,\rhat + V(r)\,\bnabla_h Y^m_\ell$, where $k = (\ell, m, n)$ denotes a specific mode, and $\bnabla_h$ is the horizontal covariant derivative \citep[e.g.][]{lavely92, jcd_notes}. Rewriting the eigenfunction in terms of generalised spherical harmonics and simplifying (see appendix~\ref{symbs}, Eq.~[\ref{recurse}] and Eq.~[\ref{specuse}]), we obtain expressions for the three components $(r,\theta,\phi)$ of the eigenfunction,
\begin{eqnarray}
\xi_{k,r} &=& \gamma_\ell\,U(r)\,Y^{0m}_\ell,\nonumber\\
\xi_{k,\theta} = \gamma_\ell\,V(r)\,\partial_\theta Y^{0m}_\ell &=& \frac{\gamma_\ell}{\sqrt2}\,V(r)\,\Omega^\ell_0\,(Y^{-1,m}_\ell - Y^{1,m}_\ell) ,\nonumber\\
\xi_{k,\phi} = \gamma_\ell\,V(r)\,\frac{1}{\sin\theta}\partial_\phi Y^{0m}_\ell &=& -i\frac{\gamma_\ell}{\sqrt2}\,V(r)\,\Omega^\ell_0\,(Y^{-1,m}_\ell + Y^{1,m}_\ell), \label{mode.eig}
\end{eqnarray}
where $U(r)$ and $V(r)$ are functions of radius alone. Using vector spherical harmonics, a general time-varying (and therefore frequency dependent) flow $\bu_0(\br,\omega)$ in a sphere may be written as
\begin{eqnarray}
\bu_0 &=& \sum_{s=0}^\infty\sum_{t=-s}^s u^t_s(r,\omega)\,Y^{t}_{s}\,\rhat + v^t_s(r,\omega)\,\bnabla_h Y^{t}_s - w^t_s(r,\omega) \,\rhat\times\bnabla_h Y^{t}_s,\nonumber\\
 &=& \sum_{s=0}^\infty\sum_{t=-s}^s \gamma_s[u^t_s\,Y^{0t}_{s}\,\rhat + v^t_s\,\bnabla_h Y^{0t}_s - w^t_s \,\rhat\times\bnabla_h Y^{0t}_s],\label{constructu}
\end{eqnarray}
where $s$ is the degree and $t$ is the azimuthal order. The first two terms capture the poloidal flows and the remaining term represents toroidal flow. In the inverse problem of determining flows, the goal is to infer the best-fit coefficients $u^t_s(r,\omega), v^t_s(r,\omega), w^t_s(r,\omega)$. 
The symmetries associated with $\bu_0$ being real in the spatio-temporal domain give us the following relationships for the coefficients,
\begin{equation}
u^{-t}_s(r,\omega) = (-1)^t [u^{t}_s(r,-\omega)]^*,\,\,\,\, v^{-t}_s(r,\omega) = (-1)^t [v^{t}_s(r,-\omega)]^*,\,\,\,\,\,w^{-t}_s(r,\omega) = (-1)^t [w^{t}_s(r,-\omega)]^*.
\end{equation}
The three components of the flow are (applying Eq.~[\ref{recurse}] and~[\ref{specuse}]),
\begin{eqnarray}
u_{r} &=& \gamma_s\,u^t_s\,Y^{0t}_{s}, \\
u_{\theta} &=& \frac{\gamma_s}{\sqrt2}\,\Omega^s_0\left[(Y^{-1t}_s - Y^{1t}_s)v^t_s - i w^t_s (Y^{-1t}_s + Y^{1t}_s) \right],\\
u_{\phi} &=& -\frac{\gamma_s}{\sqrt2}\,\Omega^s_0\left[(Y^{-1t}_s + Y^{1t}_s)i v^t_s + w^t_s (Y^{-1t}_s - Y^{1t}_s) \right].
\end{eqnarray}
Recalling equation~(\ref{connect}), we seek to write an inverse problem (also the forward problem) of the form
\begin{eqnarray}
&&\delta\langle a^{\omega+\sigma}_{k'}\,a^{\omega *}_k\rangle =H(k,k',\sigma)\sum_{s=0}^{s_{\rm max}} \sum_{t=-s}^s \int_\odot  dr\, \left[i u^t_s(r,\sigma)\, \kerne^{st}_u(r;k,k')\right.\label{invdef}\\
 &+& \left.i v^t_s(r,\sigma)\, \kerne^{st}_v(r;k,k') + w^t_s(r,\sigma)\, \kerne^{st}_w(r;k,k')\right],\nonumber
\end{eqnarray}
where the kernels $\kerne$ relate the sensitivity of the flow coefficients $u,v,$ and $w$ to the cross-spectral measurements and $H(k,k',\sigma)$ was defined in equation~(\ref{defineh}). Equation~(\ref{invdef}) states the inverse problem for cross-spectral measurements \citep[see also Eq.~35 of][]{woodard14}. 
\section{Mode coupling due to flows}
We now analyse the coupling integral for flows that occurs in equation~(\ref{connect}), i.e. where $\delta\op = -2i\omega\rho\bu_0\cdot\bnabla$ \citep[appropriate in the limit of anelastic flows,][]{gough69}, which must be evaluated in order to compute the kernels $\kerne$,
\begin{equation}
\Lambda^{k'}_k(\sigma) = - \int d\br\, \bs^*_{k'}\cdot\delta\op_\sigma\,\bs_k = 2i\omega\int_\odot\,d\br\,\rho\, \bs^*_{k'}\cdot[\bu_0(\br,\sigma)\cdot\bnabla]\bs_k, \label{coup.int}
\end{equation}
where we had defined $\Lambda^{k'}_k$ in equation~(\ref{coupledef}).  The perturbation operator for flows $\delta\op = -2i\omega\rho\bu_0(\br,\sigma)\cdot\bnabla$ is seen to depend on two frequencies, $\omega$ and $\sigma$. Since modes couple only when their frequencies are very similar (see Eq.~[\ref{couplingcon}] and subsequent discussion), we invoke the approximation $\omega\approx\nmode$, where $\nmode$ is a reference frequency close to the frequencies of the two coupled modes. Note that this approximation applies because the degree of mode coupling weakens rapidly as the difference between the two resonant frequencies increases. The latter, $\sigma$, is the temporal frequency at which the flow evolves, i.e. $\bu_0 = \bu_0(\br, \sigma)$. Anelastic flows, i.e. $\bnabla\cdot(\rho\,\bu)=0$ vary on long time scales \citep{gough69}, implying that $\sigma \ll \nmode$.
The coupling matrix $\Lambda^{k'}_k$ experiences contributions from poloidal $(u, v)$ and toroidal flows $w$. We therefore split the matrix into three components to make the computation of kernels easier,
\begin{equation}
\Lambda^{k'}_k  = \Lambda^u + \Lambda^v + \Lambda^w,\label{splittinglamb}
\end{equation}
where dependencies on $k$ and $k'$ are not explicitly noted on the right hand side.

Defining the tensor $\bT = \bnabla\bs_k$ and using the expressions from~(\ref{coordinate}) along with the definitions of mode eigenfunctions~(\ref{mode.eig}), we obtain equation~(\ref{Tsimple}).
Using recursion relations~(\ref{recurse}), we further simplify equation~(\ref{Tsimple}),
\begin{eqnarray}
T_{rr} &=&  \gamma_\ell\,\dU\,Y^{0m}_\ell,\nonumber\\
T_{r\theta} &=&   \frac{\gamma_\ell}{\sqrt2}\,\dV\,\Omega^\ell_0\,(Y^{-1m}_\ell - Y^{1m}_\ell),\nonumber\\
T_{r\phi} &=& -i\frac{\gamma_\ell}{\sqrt2}\,\dV\,\Omega^\ell_0\,(Y^{-1m}_\ell + Y^{1m}_\ell),\nonumber\\
T_{\theta r} &=& r^{-1}\gamma_\ell\,\frac{\Omega^\ell_0}{\sqrt2}\,(U - V)(Y^{-1m}_\ell - Y^{1m}_\ell),\nonumber\\
T_{\theta \theta} &=& r^{-1}{\gamma_\ell}\left[(U - \Omega^\ell_0\Omega^\ell_0V) Y^{0m}_\ell +V\Omega^\ell_0\Omega^\ell_2\frac{Y^{-2m}_\ell + Y^{2m}_\ell}{2}\right],\nonumber\\
T_{\theta \phi} &=& -ir^{-1}\frac{\gamma_\ell}{2}\,V\,\Omega^\ell_0\,\Omega^\ell_2\,(Y^{-2m}_\ell - Y^{2m}_\ell),\nonumber\\
T_{\phi r} &=& -ir^{-1}\gamma_\ell\frac{\Omega^\ell_0}{\sqrt2} (U - V) (Y^{-1m}_\ell + Y^{1m}_\ell),\nonumber\\
T_{\phi \theta} &=& ir^{-1}\frac{\gamma_\ell}{2}\,V\,\Omega^\ell_0 \Omega^\ell_{2}\left(Y^{2m}_\ell - Y^{-2m}_\ell \right),\nonumber\\
T_{\phi\phi} &=& r^{-1}\gamma_\ell\left[(U - \Omega^\ell_0\Omega^\ell_0V) Y^{0m}_\ell - V\Omega^\ell_0\Omega^\ell_{2}\frac{Y^{2m}_\ell + Y^{-2m}_\ell}{2} \right],\label{tensorT}
\end{eqnarray}
where the dot over a symbol indicates its radial derivative, i.e. $\dV = \partial_r V(r)$.

\section{Kernel for poloidal flow ($u^t_s$ coefficients)}
The algebra to compute each of these coefficients is lengthy so we limit the discussion in the main body of the text to two coefficients, $u^t_s$ and $w^t_s$. The derivation of the kernel for $v^t_s$ follows the same procedure as that for $w^t_s$. To compute $\kerne_u$, we consider radial flows in equation~(\ref{coup.int}), 
\begin{equation}
\Lambda^u = 2i\nmode\,\int_\odot d\br\, \rho \,u_r(\br,\sigma)\, (\xi^*_{k',r}\,T_{rr} + \xi^*_{k',\theta}\,T_{r\theta} + \xi^*_{k',\phi}\,T_{r\phi}),
\end{equation}
where the components $\xi_{k',}$ are described in equation~(\ref{mode.eig}). Since the eigenfunctions are real (attenuation is assumed to only affect the resonant frequency), we cease to explicitly note the conjugate symbols on the $U$ and $V$ functions (Eq.~[\ref{mode.eig}]).
{ Substituting expressions for the eigenfunction (Eq.~[\ref{mode.eig}]) and tensor $T$ (Eq.~[\ref{tensorT}]) and expanding,}
\begin{eqnarray}
&&2i\nmode\gamma_\ell\,\gamma_{\ell'}\gamma_s\int_\odot d\br\,u^t_s\, \rho\, \left[\dU\, U'\, (Y^{0m'}_{\ell'})^*\, Y^{0t}_s\,  Y^{0m}_\ell \right. \nonumber\\
&& \left. + \frac{1}{2}\,\Omega^\ell_0\,\Omega^{\ell'}_0\,\dV\, V'\, (Y^{-1m'}_{\ell'} - Y^{1m'}_{\ell'})^*\, Y^{0t}_s\,  (Y^{-1m}_\ell - Y^{1m}_\ell)\right.\nonumber\\
&&\left. + \frac{1}{2}\,\dV\,V' \Omega^\ell_0\, \Omega^{\ell'}_0\,(Y^{-1m'}_{\ell'} + Y^{1m'}_{\ell'})^*\,Y^{0t}_s\,(Y^{-1m}_\ell + Y^{1m}_\ell) \right],\nonumber \\
&=& 2i\nmode\gamma_\ell\,\gamma_{\ell'}\gamma_s\int_\odot d\br\,u^t_s\, \rho\, \left\{\dU\, U'\, (Y^{0m'}_{\ell'})^*\, Y^{0t}_s\,  Y^{0m}_\ell + \Omega^\ell_0\,\Omega^{\ell'}_0\,\dV\, V' [(Y^{1m'}_{\ell'})^*\, Y^{0t}_s\, Y^{1m}_{\ell} \right.\nonumber\\
&& \left. + (Y^{-1m'}_{\ell'} )^*\, Y^{0t}_s\,Y^{-1m}_\ell]\right\},
\end{eqnarray}
where $U',V'$ belong to the eigenfunction $\bs_{k'}$.
The Wigner 3-$j$ coefficient with $N=0$ is zero unless $\ell' + s + \ell$ is even, justifying the appearance of the $1+(-1)^{\ell'+s+\ell}$ term in the definition of the $B$ coefficient (see Appendix~\ref{symbs}).
\begin{eqnarray}
&&\int_\odot d\br\,u^t_s\,\rho\,\dU\, U'\, (Y^{0m'}_{\ell'})^*\, Y^{0t}_s\,  Y^{0m}_\ell =\nonumber\\
&& 4\pi(-1)^{m'} \begin{pmatrix} \ell' & s & \ell \\ 0 & 0 & 0\end{pmatrix} \begin{pmatrix}\ell' & s & \ell \\ -m' & t & m\end{pmatrix}\int_\odot dr\,u^t_s\,r^2\,\rho\,\dU\, U'\ \,\nonumber\\ 
&=& 4\pi(-1)^{m'} B_{\ell'sl}^{(0)+} \begin{pmatrix}\ell' & s & \ell \\ -m' & t & m\end{pmatrix}\int_\odot dr\,u^t_s\,r^2\,\rho\,\dU\, U'.
\end{eqnarray}
 Next we simplify the following terms thus
\begin{eqnarray}
&& \Omega^\ell_0\, \Omega^{\ell'}_0\,\int_\odot d\br\,u^t_s\,\rho\,\dV\, V'\, (Y^{1m'}_{\ell'})^*\, Y^{0t}_s\,  Y^{1m}_\ell \nonumber\\
 &=& -4\pi(-1)^{m'} \Omega^\ell_0\, \Omega^{\ell'}_0\begin{pmatrix} \ell' & s & \ell \\ -1 & 0 & 1\end{pmatrix} \begin{pmatrix}\ell' & s & \ell \\ -m' & t & m\end{pmatrix}\int_\odot dr\,r^2\,u^t_s\,\rho\,\dV\, V'\,,\nonumber\\
 \end{eqnarray}
and
\begin{eqnarray}
&& \Omega^\ell_0\, \Omega^{\ell'}_0\,\int_\odot d\br\,u^t_s\,\rho\,\dV\, V'\, (Y^{-1m'}_{\ell'})^*\, Y^{0t}_s\,  Y^{-1m}_\ell\nonumber\\
 & =& -4\pi(-1)^{m'} \Omega^\ell_0\, \Omega^{\ell'}_0\begin{pmatrix} \ell' & s & \ell \\ 1 & 0 & -1\end{pmatrix} \begin{pmatrix}\ell' & s & \ell \\ -m' & t & m\end{pmatrix}\int_\odot dr\,u^t_s\,r^2\,\rho\,\dV\, V'\nonumber\\
& =& -4\pi(-1)^{m'} (-1)^{\ell' + s +\ell} \Omega^\ell_0\, \Omega^{\ell'}_0\begin{pmatrix} \ell' & s & \ell \\ -1 & 0 & 1\end{pmatrix} \begin{pmatrix}\ell' & s & \ell \\ -m' & t & m\end{pmatrix}\int_\odot dr\,u^t_s\,r^2\,\rho\,\dV\, V',\nonumber\\
\label{maniple}
 \end{eqnarray}
 where in equation~(\ref{maniple}), we have use a standard manipulation of the Wigner symbol \citep[see, for instance appendix C, Section d of][]{lavely92}. The sum of these terms
gives the expression for the kernel for $u^t_s$,
\begin{equation}
\kerne^{st}_u = 8\pi \nmode\, (-1)^{m'} \gamma_{\ell'} \gamma_s \gamma_\ell\, \begin{pmatrix}\ell' & s & \ell \\ -m' & t & m\end{pmatrix} r^2\,\rho\,\left[U'\,\dU\,B_{\ell's\ell}^{(0)+}  + V'\,\dV\,B_{\ell's\ell}^{(1)+} \right],\label{kernu}
\end{equation}
and the inverse problem for the $u$ coefficients based on equations~(\ref{connect}) and~(\ref{splittinglamb}) is 
\begin{equation}
\measur^u = \sum_{s=0}^{s_{\rm max}} \sum_{t=-s}^s\int_\odot dr\,i u^t_s(r,\sigma)\, \kerne^{st}_{u}(r;k,k'),\label{invu}
\end{equation}
where $\measur^u$ is the contribution by radial flows to the cross-spectral measurement between modes $k, k'$. This is the first term of the expression~(\ref{invdef}) that we were seeking.
 
\section{Kernel for poloidal flow ($v^t_s$ coefficients)}
We do not derive the kernel here but state it for the sake of completeness. Note that it is identical to that derived by \citet{lavely92} in their equation~(C33),
\begin{eqnarray}
\kerne^{st}_v = 8\pi\rho r\nmode\gamma_{\ell'}\,\gamma_s\,\gamma_\ell\,(-1)^{m'}   \begin{pmatrix}\ell' & s & \ell \\ -m' & t & m\end{pmatrix}\left\{ (UU'-VU') B^{(1)+}_{s\ell'\ell} + \right.\nonumber\\
\left. VV'\Omega^{\ell'}_0\Omega^s_0\Omega^\ell_0\Omega^\ell_2\left[1 + (-1)^{\ell'+s+\ell}\right]\begin{pmatrix}\ell' & s & \ell \\ 1 & 1 & -2 \end{pmatrix}   
+  (UV' - \Omega^\ell_0\Omega^\ell_0VV') B^{(1)+}_{s\ell\ell'}\right\}\label{kernv}
\end{eqnarray}
The inverse problem is similar to that stated in equations~(\ref{connect}),~(\ref{splittinglamb}) and~(\ref{invu}) 
\begin{equation}
\measur^v = \sum_{s=0}^{s_{\rm max}} \sum_{t=-s}^s\int_\odot dr\, iv^t_s(r,\sigma)\,\kerne^{st}_{v}(r;k,k').\label{invv}
\end{equation}
This is the second term of the target expression~(\ref{invdef}).

\section{Kernel for toroidal flow ($w^t_s$ coefficients)}
We collect contributions from two sets of terms
\begin{eqnarray}
\Lambda^w(\sigma) = 2i\nmode\int d\br\,\rho[u_\theta(\br,\sigma)\,T_{\theta r}\, \xi^*_{k',r} + u_\theta(\br,\sigma)\,T_{\theta \theta}\, \xi^*_{k',\theta} + u_\theta(\br,\sigma)\,T_{\theta \phi}\, \xi^*_{k',\phi}]\nonumber\\
+ 2i\nmode\int d\br\,\rho[u_\phi(\br,\sigma)\,T_{\phi r}\, \xi^*_{k',r} + u_\phi(\br,\sigma)\,T_{\phi \theta} \,\xi^*_{k',\theta} + u_\phi(\br,\sigma)\,T_{\phi \phi}\, \xi^*_{k',\phi}].\label{contribw}
\end{eqnarray}
We study some of the terms here.
\subsection{Analyzing $u_\theta T_{\theta r} \xi^*_{k',r}$}
We study the first term { in the integral} in equation~(\ref{contribw}),
\begin{equation}
2\nmode\frac{\gamma_{\ell'}\gamma_s\gamma_\ell}{2}\int_\odot d\br\,\rho\,w^t_s\,r^{-1} \,(UU' - VU')\,\Omega^s_0\Omega^\ell_0 (Y^{0m'}_{\ell'})^*\, (Y^{-1t}_s + Y^{1t}_s)
(Y^{-1m}_\ell - Y^{1m}_\ell)
\end{equation}
\begin{eqnarray}
&=&-2\nmode\frac{\gamma_{\ell'}\gamma_s\gamma_\ell}{2}\,\int_\odot dr\,\rho\,w^t_s\,r^2\,4\pi(-1)^{m'} r^{-1}(UU' - VU') \Omega^s_0\Omega^\ell_0\times\nonumber\\ 
&&\left[-\begin{pmatrix}\ell' & s & \ell \\ 0 & 1 & -1 \end{pmatrix} + \begin{pmatrix}\ell' & s & \ell \\ 0 & -1 & 1 \end{pmatrix}\right]\begin{pmatrix}\ell' & s & \ell \\ -m' & t & m\end{pmatrix}\nonumber
\end{eqnarray}
\begin{eqnarray}
&=&-2\nmode\frac{\gamma_{\ell'}\gamma_s\gamma_\ell}{2}\,\int_\odot dr\,\rho\,w^t_s\,r^2\,4\pi(-1)^{m'} r^{-1}(UU' - VU') \Omega^s_0\Omega^\ell_0\times\nonumber\\
&& \left[-(-1)^{\ell'+s+\ell}\, +1\right] \begin{pmatrix}s & \ell' & \ell \\ 1 & 0 & -1 \end{pmatrix}\begin{pmatrix}\ell' & s & \ell \\ -m' & t & m\end{pmatrix}\nonumber
\end{eqnarray}
\begin{equation}
=-8\nmode\pi(-1)^{m'}\frac{\gamma_{\ell'}\gamma_s\gamma_\ell}{2}\, B^{(1)-}_{s\ell'\ell}\begin{pmatrix}\ell' & s & \ell \\ -m' & t & m\end{pmatrix}\int_\odot dr\,\rho\,w^t_s\,r (UU' - VU').
\end{equation}


\subsection{Analyzing $u_\theta T_{\theta \theta} \xi^*_{k',\theta}$}
Now we focus on the { second term in the integral in} equation~(\ref{contribw}),
\begin{eqnarray}
&&2\nmode\frac{\gamma_{\ell'}\gamma_s\gamma_\ell}{2}\Omega^{\ell'}_0\,\Omega^s_0 \int_\odot d\br\,\rho\,w^t_s\,r^{-1}\,V'(Y^{-1m'}_{\ell'} - Y^{1m'}_{\ell'})^* (Y^{-1t}_s + Y^{1t}_s)\times\nonumber\\
&&\left[(U - \Omega^\ell_0\Omega^\ell_0V) Y^{0m}_\ell +V\Omega^\ell_0\Omega^\ell_2\frac{Y^{-2m}_\ell + Y^{2m}_\ell}{2}\right]\nonumber\\
&=&2\nmode{\gamma_{\ell'}\gamma_s\gamma_\ell} \int_\odot d\br\,\rho\,w^t_s\,r^{-1}\,\left\{\frac{\Omega^{\ell'}_0\,\Omega^s_0}{2}(UV' - \Omega^\ell_0\Omega^\ell_0VV')(Y^{-1m'}_{\ell'} - Y^{1m'}_{\ell'})^* (Y^{-1t}_s + Y^{1t}_s) Y^{0m}_\ell \right.\nonumber\\
&&\left.+ VV'\Omega^{\ell'}_0\,\Omega^s_0\Omega^\ell_0\Omega^\ell_2 (Y^{-1m'}_{\ell'} - Y^{1m'}_{\ell'})^* (Y^{-1t}_s + Y^{1t}_s)\frac{Y^{-2m}_\ell + Y^{2m}_\ell}{4}\right\}\nonumber\\
&=&-2\nmode{\gamma_{\ell'}\gamma_s\gamma_\ell}\int_\odot d\br\,\rho\,w^t_s\, r^{-1}\left\{-\frac{\Omega^{\ell'}_0\,\Omega^s_0}{2}(UV' - \Omega^\ell_0\Omega^\ell_0VV')(Y^{-1m'}_{\ell'} - Y^{1m'}_{\ell'})^* (Y^{-1t}_s + Y^{1t}_s) Y^{0m}_\ell \right.\nonumber\\
&&\left.- \frac{VV'}{4}\Omega^{\ell'}_0\,\Omega^s_0\Omega^\ell_0\Omega^\ell_2 [(Y^{-1m'}_{\ell'})^* Y^{1t}_s Y^{-2m}_\ell - (Y^{1m'}_{\ell'})^* Y^{-1t}_s Y^{2m}_\ell]\right\}\nonumber\\
&=&-8\pi\nmode(-1)^{m'}{\gamma_{\ell'}\gamma_s\gamma_\ell}\int_\odot dr\,\rho\,w^t_s\,r^2\, r^{-1}\left\{\frac{\Omega^{\ell'}_0\,\Omega^s_0}{2}(UV' - \Omega^\ell_0\Omega^\ell_0VV')\left[\begin{pmatrix}\ell' & s & \ell \\ 1 & -1 & 0 \end{pmatrix} - \begin{pmatrix}\ell' & s & \ell \\ -1 & 1 & 0 \end{pmatrix} \right] \right.
\nonumber\\
&&\left. + \frac{VV'}{4}\Omega^{\ell'}_0\,\Omega^s_0\Omega^\ell_0\Omega^\ell_2 \left[ \begin{pmatrix}\ell' & s & \ell \\ 1 & 1 & -2 \end{pmatrix} - \begin{pmatrix}\ell' & s & \ell \\ -1 & -1 & 2 \end{pmatrix}\right] \right\}\begin{pmatrix}\ell' & s & \ell \\ -m' & t & m\end{pmatrix}\nonumber\\
&=& -8\pi\nmode(-1)^{m'}{\gamma_{\ell'}\gamma_s\gamma_\ell}\int_\odot dr\,\rho\,w^t_s\,r^2\,r^{-1}\left\{\frac{\Omega^{\ell'}_0\,\Omega^s_0}{2}\left[1 - (-1)^{\ell'+s+\ell} \right] \begin{pmatrix}s & \ell & \ell' \\ -1 & 0 & 1 \end{pmatrix}(UV' - \Omega^\ell_0\Omega^\ell_0VV') \right.
\nonumber\\
&&\left. + \frac{VV'}{4}\Omega^{\ell'}_0\,\Omega^s_0\Omega^\ell_0\Omega^\ell_2 \left[ 1- (-1)^{\ell'+s+\ell}\right] \begin{pmatrix}\ell' & s & \ell \\ 1 & 1 & -2 \end{pmatrix} \right\}\begin{pmatrix}\ell' & s & \ell \\ -m' & t & m\end{pmatrix}\nonumber\\
&=& 8\pi\nmode(-1)^{m'}{\gamma_{\ell'}\gamma_s\gamma_\ell}\begin{pmatrix}\ell' & s & \ell \\ -m' & t & m\end{pmatrix}\int_\odot dr\,\rho\,w^t_s\,r\left\{\frac{1}{2} B^{(1)-}_{s\ell\ell'}(UV' - \Omega^\ell_0\Omega^\ell_0VV') \right.
\nonumber\\
&&\left. - \frac{VV'}{4}\Omega^{\ell'}_0\,\Omega^s_0\Omega^\ell_0\Omega^\ell_2 \left[ 1- (-1)^{\ell'+s+\ell}\right] \begin{pmatrix}\ell' & s & \ell \\ 1 & 1 & -2 \end{pmatrix} \right\}.\label{uthtth}
\end{eqnarray}

\subsection{Overall sum}
Albeit tedious, this analysis may be applied to the remaining terms in equation~(\ref{contribw}).
Summing all terms together, we obtain
\begin{eqnarray}
&&\Lambda^w = 8\pi\nmode\gamma_{\ell'}\gamma_s\gamma_\ell(-1)^{m'} \int_\odot dr\,\rho\,w^t_s\,r\,\left\{-(UU' - VU') B^{(1)-}_{s\ell'\ell} + (UV' - \Omega^\ell_0\Omega^\ell_0VV')B^{(1)-}_{s\ell\ell'} \right.\nonumber\\
&&\left. - {VV'}{}\Omega^{\ell'}_0\,\Omega^s_0\Omega^\ell_0\Omega^\ell_2 \left[ 1- (-1)^{\ell'+s+\ell}\right] \begin{pmatrix}\ell' & s & \ell \\ 1 & 1 & -2 \end{pmatrix} \right\}\begin{pmatrix}\ell' & s & \ell \\ -m' & t & m\end{pmatrix}.
\end{eqnarray}
The kernel for the $w$ coefficients is given by
\begin{eqnarray}
&&\kerne^{st}_w = 8\pi\nmode\gamma_{\ell'}\gamma_s\gamma_\ell(-1)^{m'} \rho\,r\,\left\{-(UU' - VU') B^{(1)-}_{s\ell'\ell} + (UV' - \Omega^\ell_0\Omega^\ell_0VV')B^{(1)-}_{s\ell\ell'} \right.\nonumber\\
&&\left. - {VV'}{}\Omega^{\ell'}_0\,\Omega^s_0\Omega^\ell_0\Omega^\ell_2 \left[ 1- (-1)^{\ell'+s+\ell}\right] \begin{pmatrix}\ell' & s & \ell \\ 1 & 1 & -2 \end{pmatrix} \right\}\begin{pmatrix}\ell' & s & \ell \\ -m' & t & m\end{pmatrix},\label{kernelw}
\end{eqnarray}
and the inverse problem is similar to that stated in equations~(\ref{connect}),~(\ref{splittinglamb}),~(\ref{invu}) and~(\ref{invv})
\begin{equation}
\measur^w(\sigma) = \sum_{s=0}^{s_{\rm max}} \sum_{t=-s}^s\int_\odot dr\, w^t_s(r,\sigma)\,\kerne^{st}_{w}(r;k,k').\label{invw}
\end{equation}
Note the difference in sign in the $UU'-VU'$ term for the $w^t_s$ kernel given in equation~(\ref{kernelw}) and equation~(C34) of \citet{lavely92} - the latter appear to have made typographical errors. Equation~(\ref{invw}) is the third term of the inverse problem that was posed in~(\ref{invdef}).

\section{Kernels for anelastic flows}
The anelastic condition on flows \citep{gough69}, $\bnabla\cdot[\rho\bu_0(\br,\sigma)] = 0$ applies to small Mach-number flows in stratified environments, relevant to convective flow in the solar interior (away from the surface layers).
In principle, a time varying flow will cause the continuity equation to possess a time derivative of density as well. We ignore this on the basis that seismic signals are minimally influenced by density changes. The solenoidal condition on $\rho\bu_0$ has the following consequences
\begin{eqnarray}
&&\bnabla\cdot (\rho \bT^t_s) = -\bnabla\cdot(\rho(r) w^t_s(r)\, \rhat\times\bnabla_h Y^t_s) \nonumber\\
&&= -\frac{1}{r^2}\partial_r (r^2\rho w^t_s)\rhat\cdot (\rhat\times\bnabla_h Y^t_s) - \rho\,w^t_s(r)\bnabla_h\cdot(\rhat\times\bnabla_h Y^t_s) = 0,\nonumber\\
&&\bnabla\cdot(\rho\, {\bf P}^t_s) = \bnabla\cdot (\rho\,u^t_s(r) Y^t_s \rhat +\rho\, v^t_s(r) \bnabla_h Y^t_s) = 
\left[\frac{1}{r^2}\partial_r (r^2\rho\, u^t_s) - \frac{s(s+1)}{r}\rho\, v^t_s\right]Y^t_s = 0\label{divfree},
\end{eqnarray}
{ where $\bT^t_s$ and ${\bf P}^t_s$ are the toroidal and poloidal components of the flow.}

Consider the kernel for the $u^t_s$ coefficients, $\kerne_u$ (Eq.~[\ref{kernu}]). Introduce the symbol $\curlF$ that includes all factors not dependent on $r$ such that
\begin{equation}
\kerne^{st}_u(r) =  \rho r^2 u^t_s (U'\dU B^{(0)+}_{\ell's\ell} + V'\dV B^{(1)+}_{\ell's\ell})\curlF.
\end{equation}
Now the inverse problem, given in equation~(\ref{invu}) is
\begin{eqnarray}
i\curlF\int_\odot dr\,\rho r^2 u^t_s (U'\dU B^{(0)+}_{\ell's\ell} + V'\dV B^{(1)+}_{\ell's\ell}) = \nonumber\\
i\curlF\int_\odot dr\,\frac{\rho r^2 u^t_s}{2} [(U'\dU + U\dU') B^{(0)+}_{\ell's\ell} + (V'\dV + V\dV') B^{(1)+}_{\ell's\ell}]\label{sym}\\
+\,i\curlF\int_\odot dr\,\frac{\rho r^2 u^t_s}{2} [(U'\dU - U\dU') B^{(0)+}_{\ell's\ell} + (V'\dV - V\dV') B^{(1)+}_{\ell's\ell}].
\label{antisym}
\end{eqnarray}
The term~(\ref{sym}) is symmetric between the primed and non-primed eigenfunctions whereas term~(\ref{antisym}) is anti-symmetric. However, the presence of the $i$ pre-multiplying both terms indicates that the terms are Hermitian~(\ref{antisym}) and anti-Hermitian~(\ref{sym}) respectively. 
The modified anelatic-flow kernel for $u$, denoted by $\tilde\kerne^{st}_u$, will only contain the Hermitian term, and is given by
\begin{equation}
\tilde \kerne^{st}_u = 8\pi\nmode\rho r^2 (-1)^{m'} \gamma_{\ell'} \gamma_s \gamma_\ell\, \begin{pmatrix}\ell' & s & \ell \\ -m' & t & m\end{pmatrix}  \left(\frac{U'\dU - U\dU'}{2} B^{(0)+}_{\ell's\ell} + \frac{V'\dV - V\dV'}{2} B^{(1)+}_{\ell's\ell}\right).\label{anu}
\end{equation}
Equation~(\ref{anu}), the anelastic-flow kernel for $u$, is identical to the expression (D18) in \citet{lavely92}.
The anti-Hermitian term is removed through integration by parts,
\begin{eqnarray}
i\curlF\int_\odot dr\,\frac{\rho r^2 u^t_s}{2} [(U'\dU + U\dU') B^{(0)+}_{\ell's\ell} + (V'\dV + V\dV') B^{(1)+}_{\ell's\ell}] = \nonumber\\
i\curlF\int_\odot dr\,\left[\partial_r\left(\rho r^2 u^t_s\frac{U'U B^{(0)+}_{\ell's\ell} + {V'V} B^{(1)+}_{\ell's\ell}}{2}\right) - \frac{U'U B^{(0)+}_{\ell's\ell} + {V'V} B^{(1)+}_{\ell's\ell}}{2}\,\partial_r(\rho r^2 u^t_s)\right]\\
=-i\curlF \frac{s(s+1)}{2} \int_\odot dr\, \rho r v^t_s (U'U B^{(0)+}_{\ell's\ell} + {V'V} B^{(1)+}_{\ell's\ell}),\label{correction}
\end{eqnarray}
where we have assumed that $u^t_s$ is zero on the boundaries (causing the total derivative term to drop out), and where we invoke the divergence-free condition~(\ref{divfree}).
%
Therefore the symmetric term, i.e. equation~(\ref{sym}), contributes to the kernel for the $v$-coefficients. Now to correct $\kerne^{st}_v$ to account for term~(\ref{correction}), we first introduce the following relationships \citep[see Appendix D, Eq. D7 of][]{lavely92},
\begin{eqnarray}
&&[s(s + 1)(\ell-1 ) (\ell + 2)]^{1/2}\begin{pmatrix}\ell' & s & \ell \\ 1 & 1 & -2 \end{pmatrix} + 
[s(s + 1)\ell(\ell +1) ]^{1/2}\begin{pmatrix}\ell' & s & \ell \\ 1 & -1 & 0 \end{pmatrix}=\nonumber\\
&&\{ \ell'(\ell' +1) - s(s + 1) - \ell(\ell +1)\}\begin{pmatrix}\ell' & s & \ell \\ 1 & 0 & -1 \end{pmatrix},\\
&&\Omega^s_0\,\Omega^\ell_2\begin{pmatrix}\ell' & s & \ell \\ 1 & 1 & -2 \end{pmatrix} + 
\Omega^s_0\,\Omega^\ell_0\begin{pmatrix}\ell' & s & \ell \\ 1 & -1 & 0 \end{pmatrix}=\nonumber\\
&&\frac{1}{2}\{ \ell'(\ell' +1) - s(s + 1) - \ell(\ell +1)\}\begin{pmatrix}\ell' & s & \ell \\ 1 & 0 & -1 \end{pmatrix}. \nonumber
\end{eqnarray}
Consider the definition of $B^{(1)+}_{s\ell\ell'}$ (Appendix~\ref{symbs}),
\begin{eqnarray}
B^{(1)+}_{s\ell\ell'} = -(1 + (-1)^{s+\ell+\ell'})\Omega^s_0\Omega^{\ell'}_0 \begin{pmatrix}s & \ell & \ell' \\ -1 & 0 & 1 \end{pmatrix} = -(1 + (-1)^{s+\ell+\ell'})\Omega^s_0\Omega^{\ell'}_0 \begin{pmatrix}\ell' & s & \ell \\ 1 & -1 & 0 \end{pmatrix}. \nonumber
\end{eqnarray}
In the equation for $\kerne_v$ in equation~(\ref{kernv}), we manipulate the pair of terms
\begin{eqnarray}
-V'V\left[\Omega^\ell_0\,\Omega^\ell_0\,B^{(1)+}_{s\ell\ell'} - (1 + (-1)^{s+\ell+\ell'})\Omega^s_0\,\Omega^\ell_2\,\Omega^\ell_0\,\Omega^{\ell'}_0\,\begin{pmatrix}\ell' & s & \ell \\ 1 & 1 & -2 \end{pmatrix}\right]=\nonumber\\
V'V\Omega^\ell_0\,\Omega^{\ell'}_0\,(1 + (-1)^{s+\ell+\ell'})\left[\Omega^\ell_0 \Omega^s_0\begin{pmatrix}\ell' & s & \ell\\ 1 & -1 & 0 \end{pmatrix} + \Omega^s_0\,\Omega^\ell_2\,\begin{pmatrix}\ell' & s & \ell \\ 1 & 1 & -2 \end{pmatrix}\right],\nonumber\\
\frac{V'V}{2}\{ \ell'(\ell' +1) - s(s + 1) - \ell(\ell +1)\}\Omega^\ell_0\,\Omega^{\ell'}_0\,(1 + (-1)^{s+\ell+\ell'}) \begin{pmatrix}\ell' & s & \ell \\ 1 & 0 & -1 \end{pmatrix} = \nonumber\\
-\frac{V'V}{2}\{ \ell'(\ell' +1) - s(s + 1) - \ell(\ell +1)\} B^{(1)+}_{\ell' s\ell} = \frac{V'V}{2}\{ s(s + 1) + \ell(\ell +1)-\ell'(\ell' +1)\} B^{(1)+}_{\ell' s\ell}\label{simpvv}.
\end{eqnarray}
%
%
Substituting expression~(\ref{simpvv}) into equation~(\ref{kernv}), and adding term~(\ref{correction}), we obtain an expression for the anelastic form of the kernel for $v$, denoted by $\tilde\kerne_v$, 
\begin{eqnarray}
\tilde\kerne_v = 8\pi\rho r\nmode\gamma_{\ell'}\,\gamma_s\,\gamma_\ell\,(-1)^{m'}   \begin{pmatrix}\ell' & s & \ell \\ -m' & t & m\end{pmatrix}\left\{ UU' B^{(1)+}_{s\ell'\ell}  -\frac{s(s+1)}{2}(U'U B^{(0)+}_{\ell's\ell} + {V'V} B^{(1)+}_{\ell's\ell}) \right.\nonumber\\
\left. +  \frac{V'V}{2}[ s(s + 1) + \ell(\ell +1)-\ell'(\ell' +1)] B^{(1)+}_{\ell' s\ell} + UV'B^{(1)+}_{s\ell\ell'} -VU' B^{(1)+}_{s\ell'\ell}\right\}.\label{anvee1}
\end{eqnarray}
Applying the relationship $B^{(1)+}_{s\ell'\ell} = \frac{1}{2}[ s(s + 1) + \ell(\ell +1)-\ell'(\ell' +1)] B^{(0)+}_{s\ell' \ell}$ and $B^{(0)+}_{s\ell' \ell} = B^{(0)+}_{\ell's \ell}$ \citep[Eq. C45 and C46 of][]{lavely92}, we simplify equation~(\ref{anvee1}) further to obtain the anelastic kernel for $v$,
\begin{eqnarray}
&&\tilde\kerne^{st}_v = 8\pi\rho r\nmode\gamma_{\ell'}\,\gamma_s\,\gamma_\ell\,(-1)^{m'}   \begin{pmatrix}\ell' & s & \ell \\ -m' & t & m\end{pmatrix}\times\nonumber\\
&&\left\{\frac{UU' B^{(0)+}_{\ell's\ell} + V'VB^{(1)+}_{\ell' s\ell}}{2}[ \ell(\ell +1)-\ell'(\ell' +1)]  + UV'B^{(1)+}_{s\ell\ell'} -VU' B^{(1)+}_{s\ell'\ell}\right\}.\label{anvee}
\end{eqnarray}
Equation~(\ref{anvee}), the anelastic-flow kernel for $v$, is identical to the expression (D19) in \citet{lavely92}. Note that $B^{(1)+}_{s\ell\ell'} = B^{(1)+}_{\ell'\ell s}$ and $B^{(1)+}_{s\ell'\ell} = B^{(1)+}_{\ell\ell' s}$ always hold.

\begin{figure}
\begin{centering}
\includegraphics[width=\linewidth]{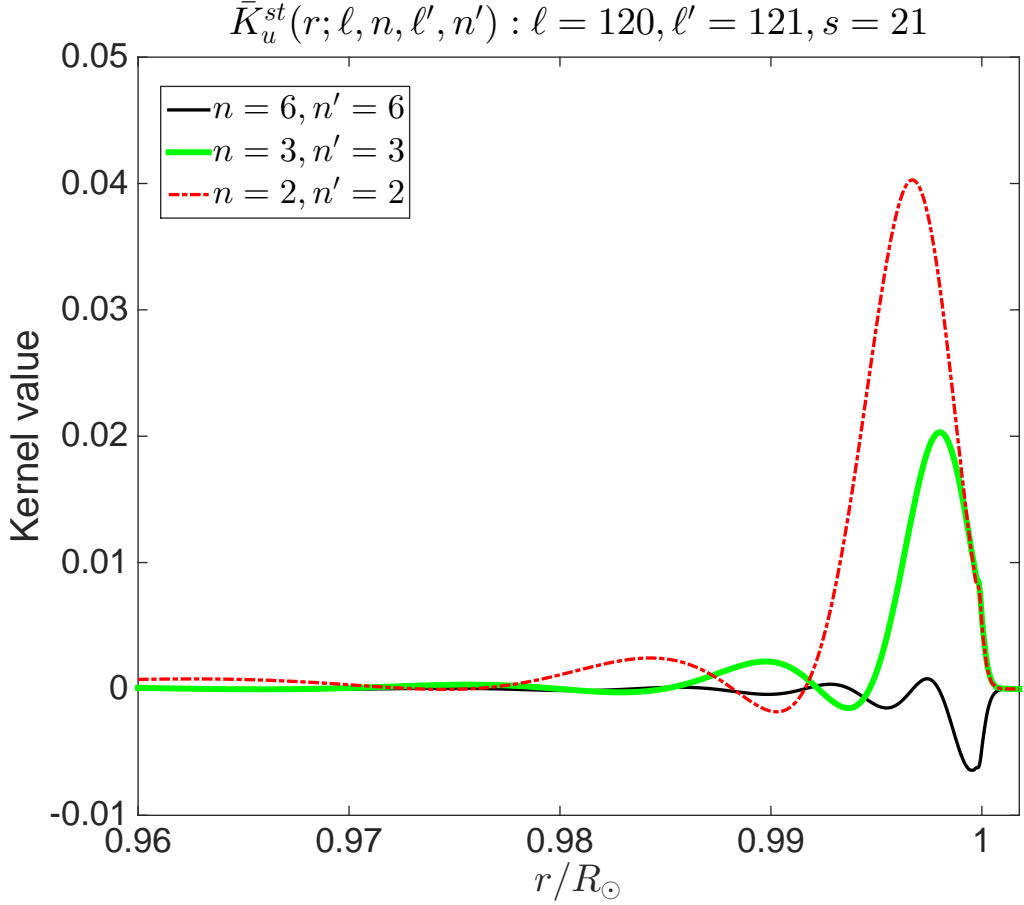}
\caption{Examples of continuity-constrained $\bar\kerne_u(r)$ (Eq.~[\ref{finalukern}]) to interpret cross-spectral measurements between a variety of radial orders $n,n'$ and $\ell = 120$, $\ell' = 121$ and $s = 21$. The $B^{(0)+}$ and $B^{(1)+}$ terms in the expression for the kernel impliy that it is non-zero only for even $\ell + s +\ell'$ and for $(\ell,n)\neq(\ell',n')$. A Wigner-3$j$ symbol in the expression for the kernel in equation~(\ref{finalukern}), which codifies the dependencies on $m, m'$ and $t$, serves only to modulate the amplitude and overall sign of the kernel. Thus we ignore it (by setting it to 1) here.
\label{kernuexamp}}
\end{centering}
\end{figure}

Next, we manipulate the terms involving $VV'$ in the expression for $\kerne_w$ in equation~(\ref{kernelw}),
\begin{eqnarray}
&&V'V\left[\Omega^\ell_0\,\Omega^\ell_0\,B^{(1)-}_{s\ell\ell'} + (1 - (-1)^{s+\ell+\ell'})\Omega^s_0\,\Omega^\ell_2\,\Omega^\ell_0\,\Omega^{\ell'}_0\,\begin{pmatrix}\ell' & s & \ell \\ 1 & 1 & -2 \end{pmatrix}\right].
\end{eqnarray}
We use the identities $B^{(1)-}_{\ell's\ell} = B^{(1)-}_{\ell'\ell s} = B^{(1)-}_{s\ell\ell'}$ \citep[e.g., Eq.~46 of][and references therein]{lavely92} and
\begin{eqnarray}
(1-(-1)^{\ell'+s+\ell})\Omega^s_0 \Omega^\ell_2 \Omega^\ell_0 \Omega^{\ell'}_0 \begin{pmatrix}\ell' & s & \ell \\ 1 & 1 & -2 \end{pmatrix} - \Omega^\ell_0\Omega^\ell_0 B^{(1)-}_{s\ell\ell'} = \nonumber\\
\frac{1}{2} [\ell'(\ell' +1) - \ell(\ell +1) - s(s + 1)] B^{(1)-}_{\ell's\ell},
\end{eqnarray}
we obtain the following relationship
\begin{eqnarray}
(1-(-1)^{\ell'+s+\ell})\Omega^s_0 \Omega^\ell_2 \Omega^\ell_0 \Omega^{\ell'}_0 \begin{pmatrix}\ell' & s & \ell \\ 1 & 1 & -2 \end{pmatrix} + \Omega^\ell_0\Omega^\ell_0 B^{(1)-}_{s\ell\ell'} = \nonumber\\
\frac{1}{2} [\ell'(\ell' +1) + \ell(\ell +1) - s(s + 1)] B^{(1)-}_{\ell's\ell}.
\end{eqnarray}
Rearranging equating~(\ref{kernelw}), we obtain a simplified expression for the kernel for the $w$-coefficients, denoted as $\tilde\kerne^{st}_w$,
\begin{eqnarray}
&&\tilde\kerne^{st}_w = 8\pi\nmode\gamma_{\ell'}\gamma_s\gamma_\ell(-1)^{m'} \rho\,r\,\left\{-UU' + VU' + UV' \right.\nonumber\\
&&\left.-\frac{VV'}{2}[\ell'(\ell' +1) + \ell(\ell +1) - s(s + 1)]\right\}B^{(1)-}_{\ell's\ell} \begin{pmatrix}\ell' & s & \ell \\ -m' & t & m\end{pmatrix}.\label{anw}
\end{eqnarray}
Equations~(\ref{anw}) and~(\ref{kernelw}) are necessarily identical since { toroidal} flows automatically satisfy continuity. The anelastic kernel~(\ref{anw}) is identical to expression~(D20) of \citet{lavely92} in two places; indeed the typographical error appears to not have carried forward from their equation~(C34). 
Finally, noting again from the continuity equation that
\begin{equation}
\rho v^t_s = \frac{r\partial_r (\rho u^t_s) + 2\rho u^t_s}{s(s+1)},\label{expvst}
\end{equation}
we substitute expression~(\ref{expvst}) for $v^t_s$ to obtain the following constrained inverse problem for anelastic flows that has only two scalar parameters,
\begin{equation}
\delta\langle a^{\omega+\sigma}_{k'}\,a^{\omega *}_k\rangle =H(k,k',\sigma)\sum_{s=0}^{s_{\rm max}} \sum_{t=-s}^s\int_\odot  dr\, i u^t_s(r)\, \bar\kerne^{st}_u(r;k,k') + w^t_s(r)\, \tilde\kerne^{st}_w(r;k,k'),\label{aninvdef}
\end{equation}
where we assume that the function $u^t_s\,\tilde\kerne_v$ vanishes at the boundary and
\begin{equation}
\bar\kerne^{st}_u = \tilde\kerne^{st}_u + \frac{ \partial_r(r\ln\rho)\, \tilde\kerne^{st}_v - r\partial_r\tilde\kerne^{st}_v}{s(s+1)}.\label{finalukern}
\end{equation}
Note that our formulation of the inverse problem differs from that of \citet{lavely92} in that we only invert for two scalar functions, one a poloidal component $u^t_s$ and the other the toroidal, $w^t_s$. From $u^t_s$, we obtain the other poloidal component $v^t_s$ using the continuity equation, i.e. specifically equation~(\ref{expvst}). Computing Wigner-3$j$ symbols for high degrees and orders (i.e. $\ell > 80$ for instance) represents a non-trivial challenge. We use an implementation of an algorithm described in \citet{schulten1975} downloaded from the SLATEC library for these calculations.
In Figures~\ref{kernuexamp} and~\ref{kernwexamp}, we show plots of continuity-constrained kernels for $u^t_s$ and $w^t_s$ respectively. 

\begin{figure}
\begin{centering}
\includegraphics[width=\linewidth]{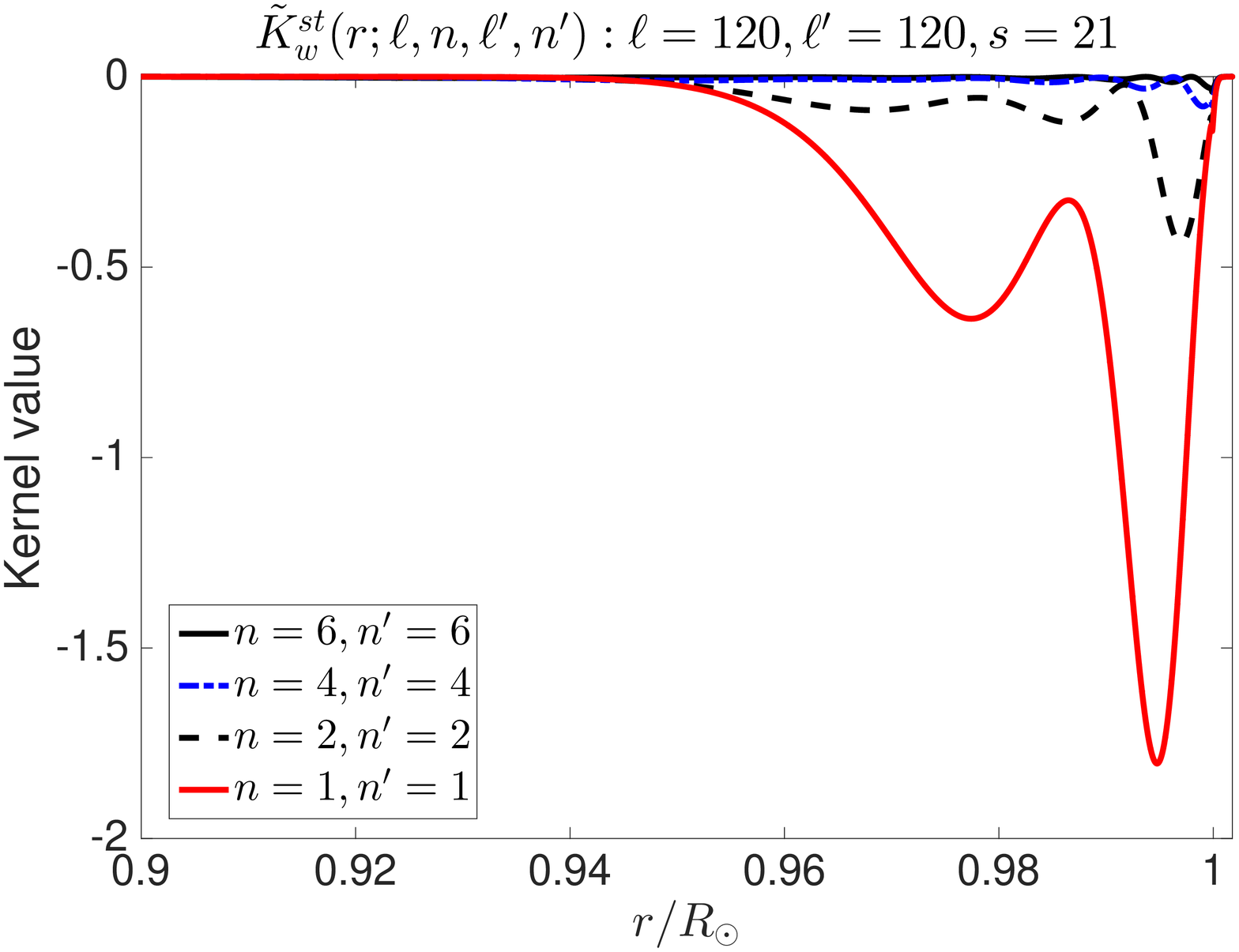}
\caption{Examples of self-coupled $\tilde\kerne_w(r)$ (Eq.~[\ref{anw}]) with $s = 21$, $\ell = 120 = \ell'$ and $n' = n$. The  $B^{(1)-}$ term in the expression for the kernel implies that it is non-zero only for odd $\ell + s +\ell'$.  A Wigner-3$j$ symbol in the expression for the kernel in equation~(\ref{anw}), which codifies the dependencies on $m, m'$ and $t$, serves only to modulate the amplitude and overall sign of the kernel. Thus we ignore it (by setting it to 1) here. The $n=0$ or $f$-mode kernel shows no nodes in radius whereas the higher $n$ kernels have $n$ nodes in radius.
\label{kernwexamp}}
\end{centering}
\end{figure}

\begin{figure}
\begin{centering}
\includegraphics[width=\linewidth]{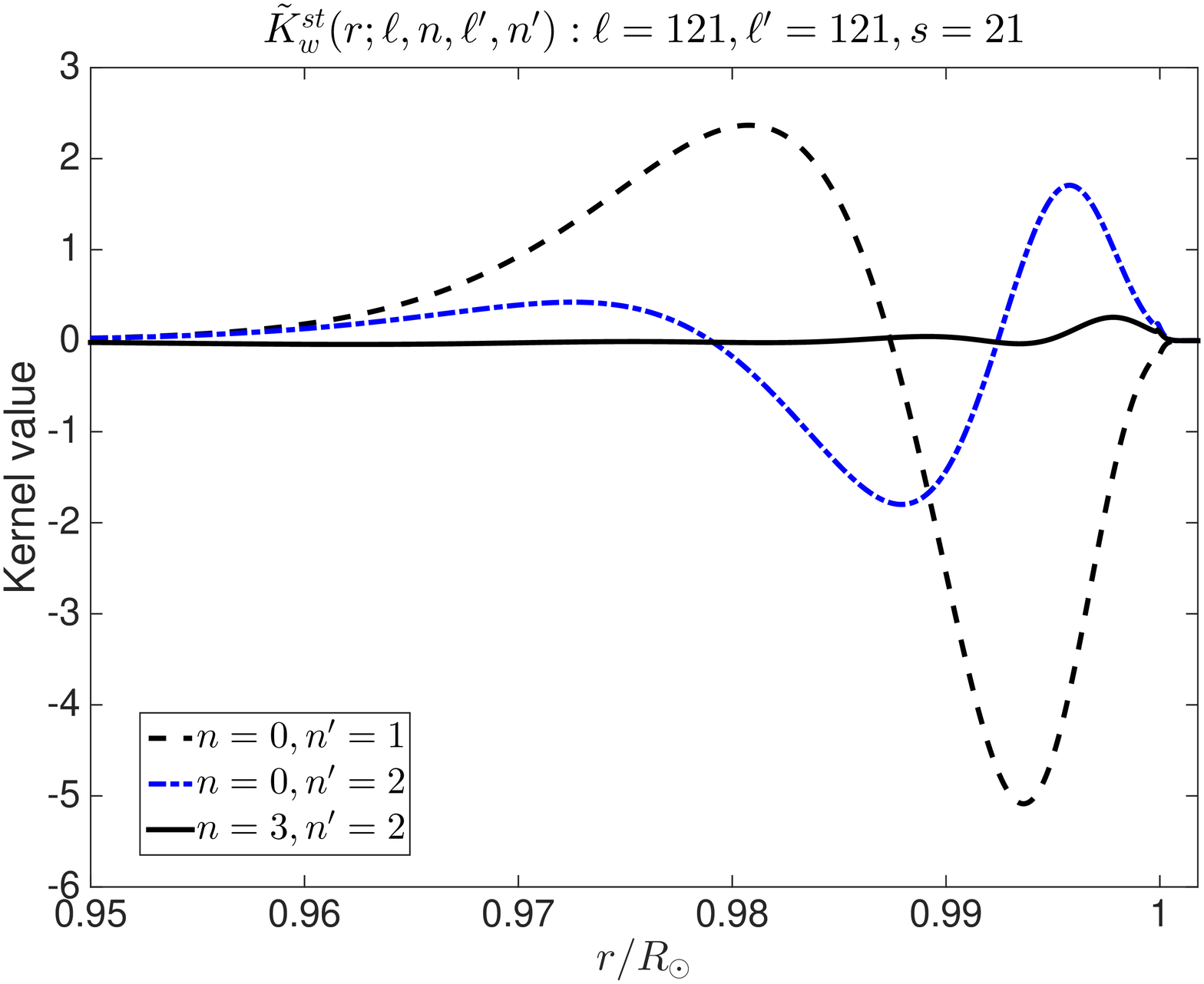}
\caption{Examples of cross-coupled $\tilde\kerne_w(r)$ (Eq.~[\ref{anw}]) with $s = 21$, $\ell = 121 = \ell'$. The number of nodes in radius of the cross-spectral kernel is equal to the maximum number of nodes of the constituent eigenfunctions.  A Wigner-3$j$ symbol in the expression for the kernel in equation~(\ref{anw}), which codifies the dependencies on $m, m'$ and $t$, serves only to modulate the amplitude and overall sign of the kernel. Thus we ignore it (by setting it to 1) here.
\label{kernwexamp}}
\end{centering}
\end{figure}

\begin{figure}
\begin{centering}
\includegraphics[width=\linewidth]{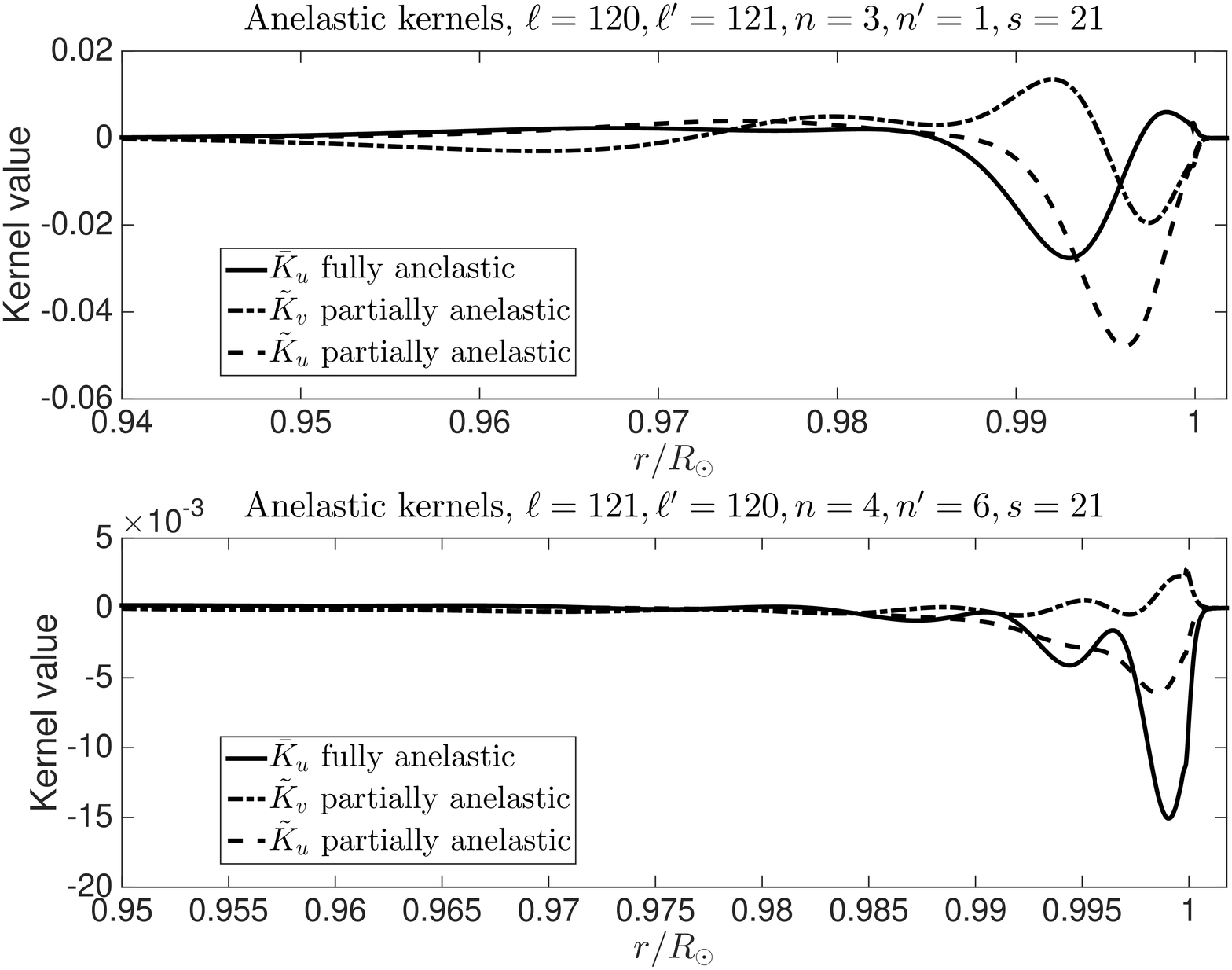}
\caption{Examples of cross-spectral coupling kernels $\tilde\kerne_u(r)$ (Eq.~[\ref{anu}]), $\tilde\kerne_v(r)$  (Eq.~[\ref{anvee}]) and $\bar\kerne_u(r)$  (Eq.~[\ref{finalukern}]) with $s = 21$. The kernels $\tilde\kerne_u$ and $\tilde\kerne_v$ have been modified to account for the continuity equation. However, \citet{lavely92} appear to have missed the issue that for anelastic flows, only one of the two expansion coefficients is required. Absorbing this condition, we obtain an inverse problem for poloidal flows that contains only one term, $\bar\kerne_u(r)$. For the cases shown here, the one-term anelastic kernel (Eq~[\ref{anu}]) is decidedly different from its constituent parts.  Wigner-3$j$ symbols in the expressions for the kernels in equations~(\ref{anu}),~(\ref{anvee}) and~(\ref{finalukern}), which codify the dependencies on $m, m'$ and $t$, serve only to modulate the amplitude and overall sign of the kernels in this plot. Thus we ignore it (by setting it to 1) here.
\label{elastic}}
\end{centering}
\end{figure}

Figure~\ref{elastic} shows kernels for anelastic poloidal flows based on equation~(\ref{finalukern}), termed ``fully anelastic" since it reduces the inference to a one-term inversion, compared with expressions~(\ref{anu}) and~(\ref{anvee}), which result in a (redundant) two-term inversion problem for $u^t_s$ and $v^t_s$ \citep[also described in][]{lavely92}.

\begin{figure}
\begin{centering}
\includegraphics[width=\linewidth]{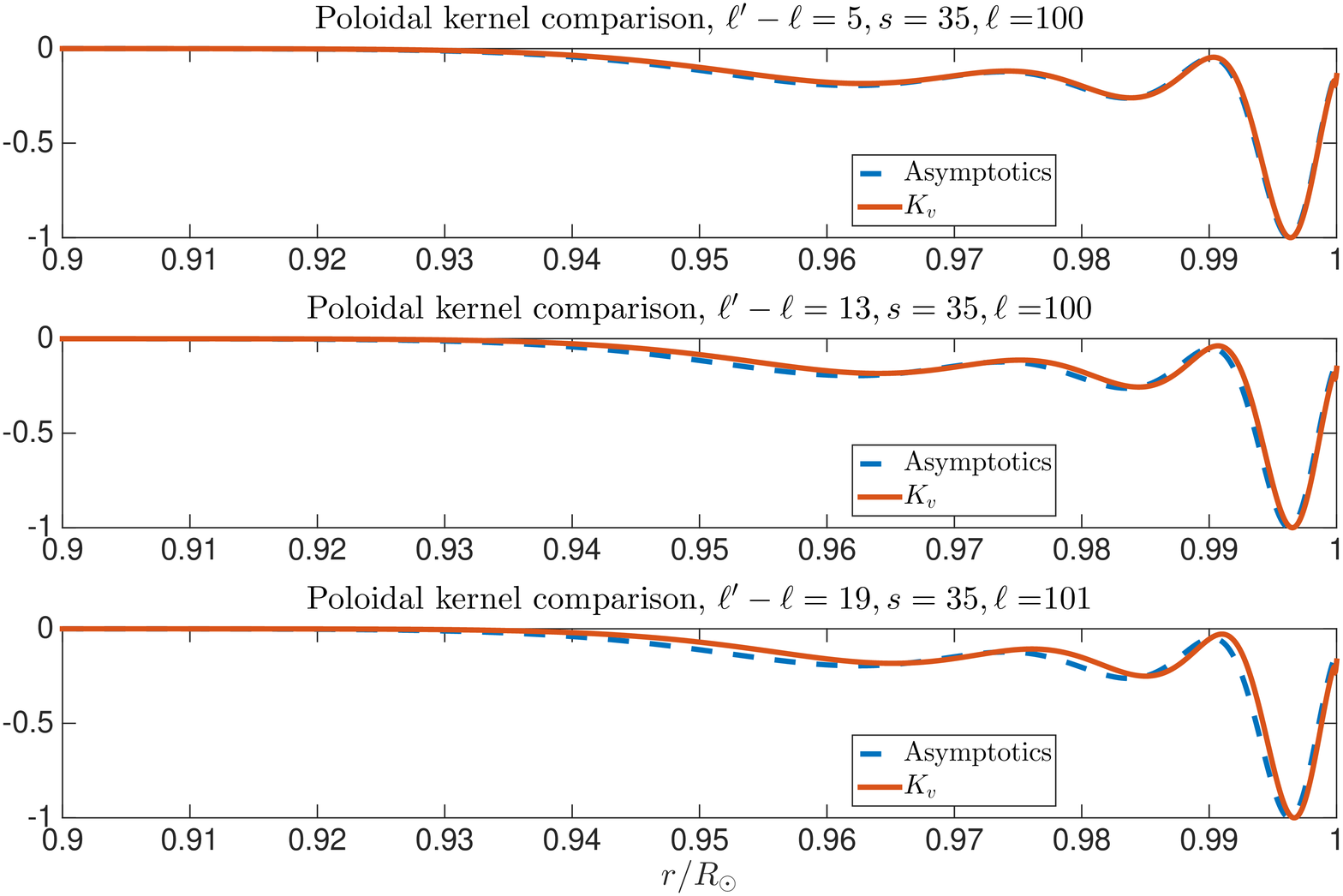}
\caption{Comparison between exact expression for the kernel for the poloidal-flow component $v^t_s$ (Eq.~[\ref{anvee}]) and asymptotics \citep[see Eq.~\ref{vas} and][]{woodard14} for radial orders $n = 2 = n'$. The kernels have been normalized by their corresponding maximum values to aid comparison. For the kernel to be non-zero, we require $|\ell' - \ell| \le s$. It is seen that when the difference between $|\ell - \ell'| \ll s$ and $s \ll \ell, \ell'$, the asymptotic expressions reproduce the exact kernel accurately. When $|\ell'-\ell|$ becomes comparable $s$, the asymptotics begin to deviate slightly from the exact kernels.  A Wigner-3$j$ symbol in the expression for the kernel in equation~(\ref{anvee}), which codifies the dependencies on $m, m'$ and $t$, serves only to modulate the amplitude and overall sign of the kernel. Thus we ignore it (by setting it to 1) here.
\label{asympv}}
\end{centering}
\end{figure}

\begin{figure}
\begin{centering}
\includegraphics[width=\linewidth]{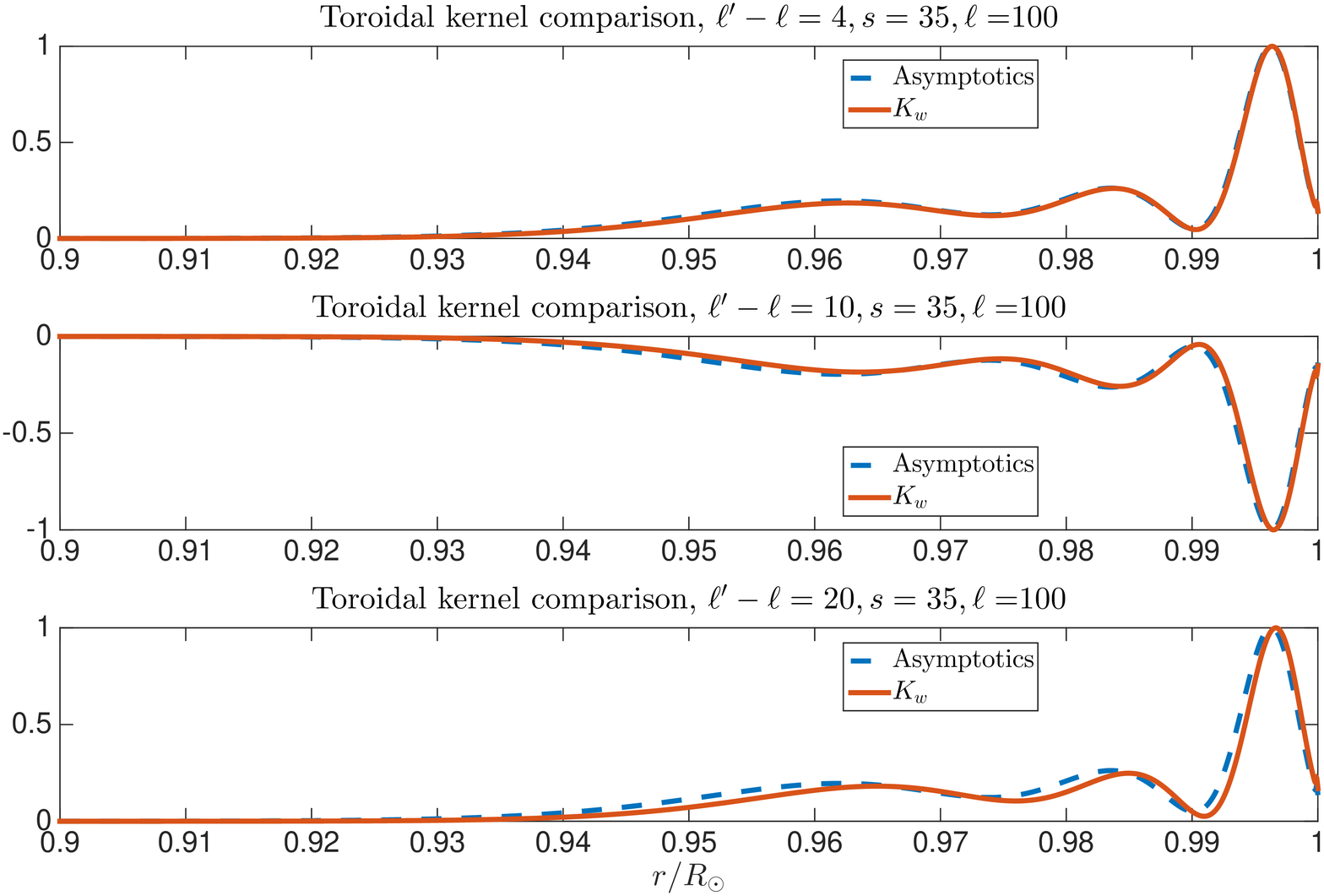}
\caption{Comparison between exact expression for the kernel for the toroidal-flow component $w^t_s$ (Eq.~[\ref{anw}]) and asymptotics \citep[see Eq.~\ref{was} and][]{woodard14} for radial orders $n = 2 = n'$. The kernels have been normalized by their corresponding maximum values to aid comparison. For the kernel to be non-zero, we require $|\ell' - \ell| \le s$. It is seen that when the difference between $|\ell - \ell'| \ll s$ and $s \ll \ell, \ell'$, the asymptotic expressions reproduce the exact kernel accurately. When $|\ell'-\ell|$ becomes comparable $s$, the asymptotics begin to deviate slightly from the exact kernels.  A Wigner-3$j$ symbol in the expression for the kernel in equation~(\ref{anw}), which codifies the dependencies on $m, m'$ and $t$, serves only to modulate the amplitude and overall sign of this plot. Thus we ignore it (by setting it to 1) here.
\label{asympw}}
\end{centering}
\end{figure}

\begin{figure}
\begin{centering}
\includegraphics[width=\linewidth]{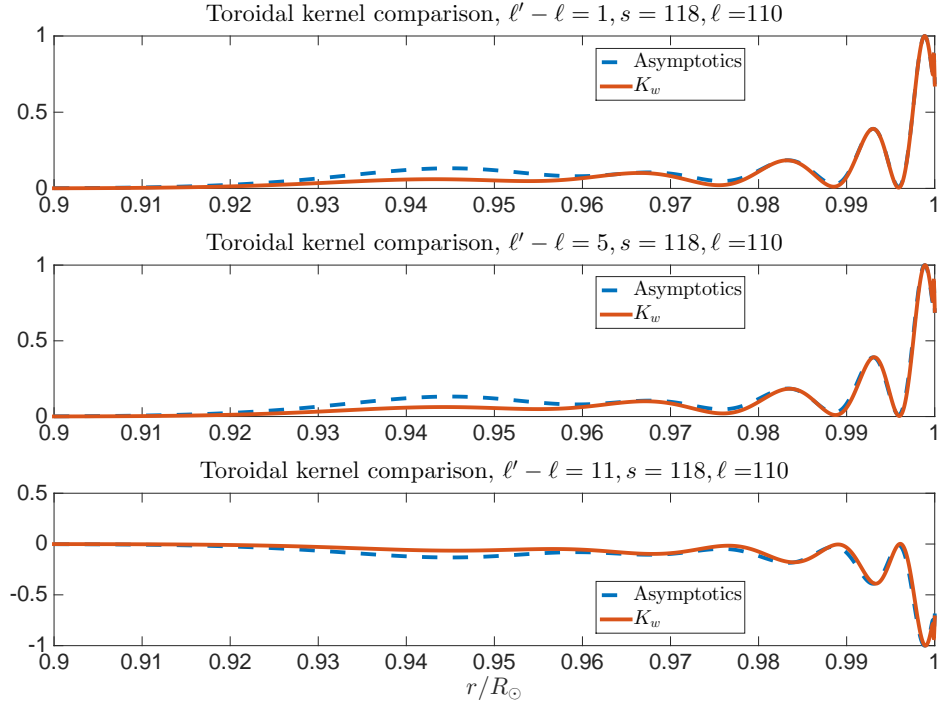}
\caption{Comparison between exact expression for the kernel for the toroidal-flow component $w^t_s$ (Eq.~[\ref{anw}]) and asymptotics \citep[see Eq.~\ref{was} and][]{woodard14} for radial orders $n = 4 = n'$. The kernels have been normalized by their corresponding maximum values to aid comparison. For the kernel to be non-zero, we require $|\ell' - \ell| \le s$. { Despite $s$ being comparable to $\ell, \ell'$, the asymptotic expressions reproduce the exact kernel relatively accurately, although there appears to be some mismatch at the lower turning point}. A Wigner-3$j$ symbol in the expression for the kernel in equation~(\ref{anw}), which codifies the dependencies on $m, m'$ and $t$, serves only to modulate the amplitude and overall sign of this plot. Thus we ignore it (by setting it to 1) here.
\label{asympw3}}
\end{centering}
\end{figure}

In Figures~\ref{asympv} through ~\ref{asympw3}, we compare the kernels for poloidal flow, computed using equation~(\ref{anvee}) and for toroidal flows using equation~(\ref{anw}), with corresponding asymptotic expressions \citep[see equations~43 through~47 of][]{woodard14}, {
\begin{equation}
K_v^{\rm asymp} = (-1)^{\ell}\,i\,r\,g(s,\ell'-\ell)\,\rho\,\ell^{\frac{3}{2}} [ U_{n\ell}^2 + \ell(\ell+1)\,V_{n\ell}^2],\label{vas}
\end{equation}
and
\begin{equation}
K_w^{\rm asymp} = (-1)^{\ell}\,r^2\,f(s,\ell'-\ell)\,\rho\,\ell^{\frac{3}{2}} [ U_{n\ell}^2 + \ell(\ell+1)\,V_{n\ell}^2],\label{was}
\end{equation}
where equations~(\ref{vas}) and~(\ref{was}) are asymptotic kernels sensitive to poloidal and toroidal flow components $v$ and $w$ respectively, $f$ and $g$ are asymptotic expressions for Wigner-3$j$ symbols in the limit $s\ll\ell, \ell'$ \citep{vorontsov11,DT98} and $|\ell' - \ell |\ll s$. These expressions also assume $|\ell - \ell'|\ll\ell$ and $n = n'$, which allow for the approximation $U\approx U'$ and $V\approx V'$.  Since only eigenfunctions for a single $(n,\ell)$ mode appear in the expression, these asymptotic kernels cannot account for cross $\ell, n$ coupling. The asymptotics match well with exact expressions even in cases where $s$ is comparable $\ell,\ell'$ and $|\ell' - \ell|$.}

\section{Kernels for Sound-speed perturbations}
\begin{figure}
\begin{centering}
\includegraphics[width=\linewidth]{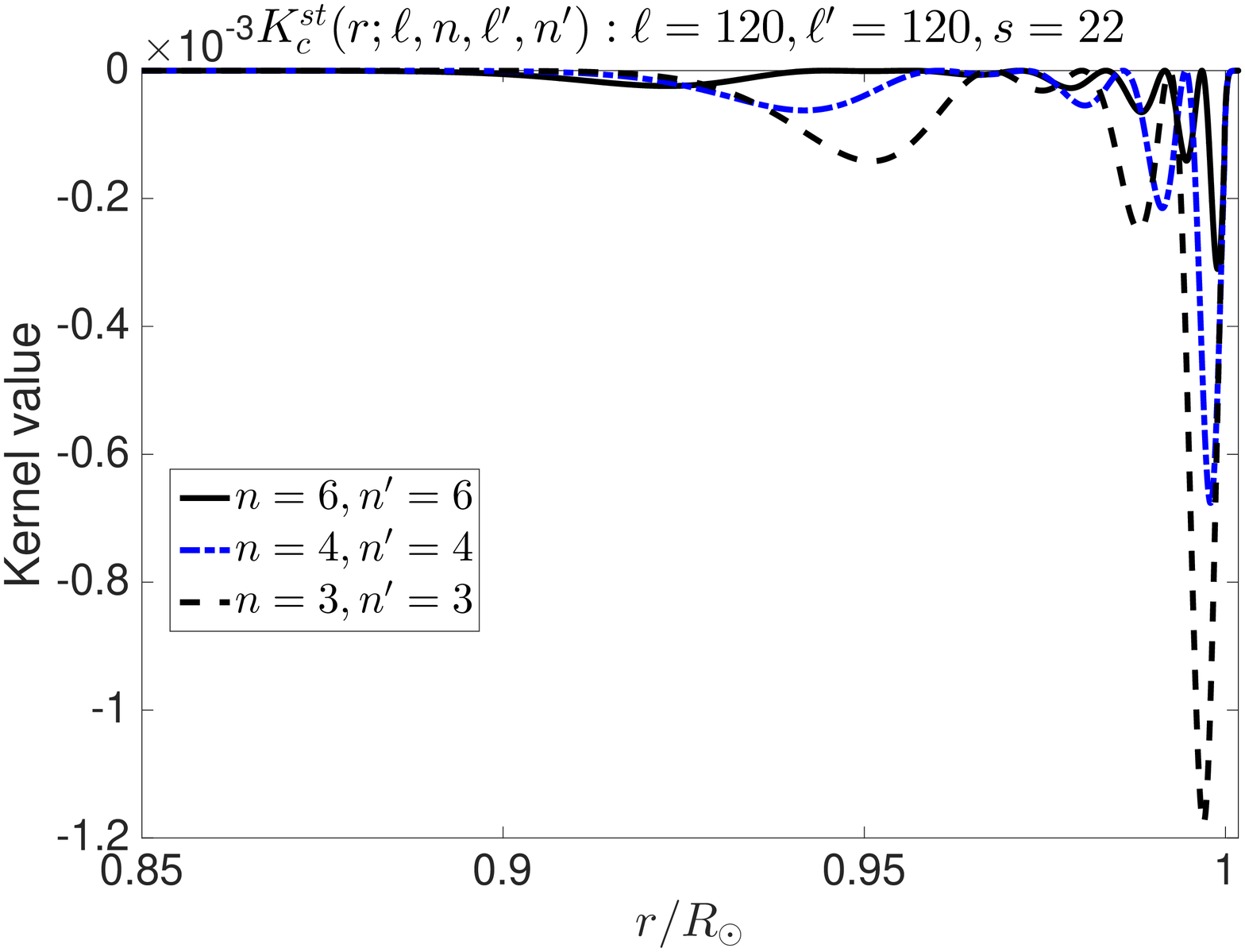}
\caption{Examples of self-coupled $\kerne_c(r)$ (Eq.~[\ref{kernc}]) with $s = 22$, $\ell = 120 = \ell'$ and $n' = n$. The  $B^{(0)+}$ term in the expression for the kernel implies that it is non-zero only for even $\ell + s +\ell'$. A Wigner-3$j$ symbol in the expression for the kernel in equation~(\ref{anw}), which codifies the dependencies on $m, m'$ and $t$, serves only to modulate the amplitude and overall sign of the kernel. Thus we ignore it (by setting it to 1) here. Primarily $p$ modes are sensitive to sound-speed perturbations. Kernels for modes of radial order $n$ have $n$ nodes in radius. 
\label{kerncexamp}}
\end{centering}
\end{figure}

Recalling the wave equation~(\ref{fullop}), sound-speed anomalies affect the wavefield through the perturbation operator $\delta\op_\omega = -\bnabla[\delta(\rho c^2)\,\bnabla\cdot]$, where $c = c(\br,\omega)$ may be time varying. The coupling integral due to this perturbation, given by equation~(\ref{coupledef}) is,
\begin{equation}
\Lambda^{k'}_k(\sigma) = \int_\odot d\br\, \bs^*_{k'}\cdot\bnabla(\delta(\rho c^2)\,\bnabla\cdot\bs_k) = -2\int_\odot d\br\, \delta(\ln c)\,\rho c^2 \bnabla\cdot\bs^*_{k'}\,\bnabla\cdot\bs_k,\label{ssp}
\end{equation}
where sound speed is positive-definite and time varying and where we have ignored the boundary terms (i.e. assuming that $\delta c = 0$ on the outer boundary). 
We now expand $\delta\ln c$ in spherical harmonics,
\begin{equation}
\delta\ln c = \sum_{s=0}^{s_{\rm max}} \sum_{t=-s}^s  c^t_s(r,\sigma)\,Y^t_s = \sum_{s=0}^{s_{\rm max}} \sum_{t=-s}^s \gamma_s\, c^t_s\,Y^{0t}_s,
\end{equation}
where because $c$ is real in the spatio-temporal domain, $c^{-t}_s(r,\sigma) = (-1)^t [c^{t}_s(r,-\sigma)]^*$, and
\begin{equation}
\bnabla\cdot\bs_k = Tr(\bT) = \left[\frac{1}{r^2}\partial_r(r^2 U) - \frac{\ell(\ell+1)}{r} V\right] Y^{m}_\ell = \left[\frac{1}{r^2}\partial_r(r^2 U) - \frac{\ell(\ell+1)}{r} V\right] \gamma_\ell\,Y^{0m}_\ell,
\end{equation}
where the tensor $\bT$ is defined in equation~(\ref{tensorT}) and the trace may be calculated as $Tr(\bT) = T_{rr} + T_{\theta\theta} + T_{\phi\phi}$. { Using equation~(\ref{tensorT}) to compute the trace, substituting it} into equation~(\ref{ssp}) and simplifying,
we obtain the following expression for the kernel for the logarithm of sound speed,
\begin{eqnarray}
\kerne^{st}_c = -8\rho\pi c^2 r^2(-1)^{m'}\gamma_\ell\,\gamma_{\ell'}\,\gamma_s\begin{pmatrix}\ell' & s & \ell \\ -m' & t & m\end{pmatrix} B^{(0)+}_{\ell's\ell}\,\left[\frac{1}{r^2}\partial_r(r^2 U') - \frac{\ell'(\ell'+1)}{r} V'\right]\times\nonumber\\
\left[\frac{1}{r^2}\partial_r(r^2 U) - \frac{\ell(\ell+1)}{r} V\right]\nonumber\\
=-8\rho\pi c^2(-1)^{m'}\gamma_\ell\,\gamma_{\ell'}\,\gamma_s\begin{pmatrix}\ell' & s & \ell \\ -m' & t & m\end{pmatrix} B^{(0)+}_{\ell's\ell}\,\left[r\dU' + 2 (U' - \Omega^{\ell'}_0\Omega^{\ell'}_0 V')\right]\left[r\dU + 2(U - \Omega^{\ell}_0\Omega^{\ell}_0 V)\right].\label{kernc}
\end{eqnarray}
Note that we may take into account time variations of sound-speed anomalies in exactly the same as we did for flows (see Eq.~[\ref{connect}]). Thus the inverse problem may be written as
\begin{equation}
\delta\langle a^{\omega+\sigma}_{k'}\,a^{\omega *}_k\rangle =H(k,k',\sigma)\sum_{s=0}^{s_{\rm max}} \sum_{t=-s}^s\int_\odot  dr\, \kerne^{st}_c\,c^t_s(r,\sigma),\label{invsee}
\end{equation}
where $c^t_s(r,\sigma)$ represents the coefficient associated with the sound-speed anomaly temporally filtered at frequency $\sigma$. We reconstruct the sound speed thus
\begin{equation}
c(\br,\sigma) = c_0(r)\,\exp\left(\sum_{s,t} c^t_s(r,\sigma)\, Y^t_s\right),\label{constructc}
\end{equation}
where $c_0(r)$ is the fiducial sound speed of the spherically symmetric solar model for which the eigenfunctions have been computed. Examples of kernels for sound-speed perturbations are shown in Figure~\ref{kerncexamp}.

\section{Discussion}
The use of cross-spectral normal-mode measurements in helioseismology represents an exciting avenue to investigate the structure of convective flows, large-scale circulations and thermal anomalies in the solar interior. A description of the measurement process applied to obtain the cross-spectra is found in e.g. \citet{woodard16}. In this article, we have derived and computed kernels for flows and sound-speed perturbations. We have also resolved minor inconsistencies in the mode-coupling theory for helioseismology set out by \citet{lavely92}. Taking into account the constraint that the continuity equation imposes on flows, we also reformulate the inverse problem for two scalar functions. To summarise, the inverse problem for continuity-preserving flows is given in equation~(\ref{aninvdef}), where the kernels and related quantities are defined in equations~(\ref{finalukern}),~(\ref{anu}),~(\ref{anvee}), and~(\ref{anw}). Other relevant symbols and a means to recompute the flow field in the spatial domain are described in equations~(\ref{expvst}),~(\ref{defineh}), and~(\ref{constructu}). The inverse problem for sound-speed perturbations is described in equation~(\ref{invsee}), and related expressions for kernels and reconstruction of the sound speed in spatial domain are described in equations~(\ref{kernc}),~(\ref{defineh}) and~(\ref{constructc}).

The kernels computed here can be used to study a variety of time-varying flow structures such as supergranulation and large-scale convection, meridional flow and associated sound-speed perturbations (should they be measurable). The formalism here can also be adopted for a starting model that is differentially rotating using perturbation theory outlined in, e.g. \citet{gough90}.  Indeed, the theory may be extended to infer interior magnetism although the theory is complicated by the fact that these measurements are sensitive not to the magnetic field directly but rather the Lorentz stress tensor. One way to proceed is to reformulate linear magnetohydrodynamics as an analogous elastodynamics problem and use geophysical literature to obtain expressions for sensitivity kernels for the fourth-rank wavespeed tensor \citep[e.g.][]{woodhouse80, DT98}.

\section*{Acknowledgments}
SMH, LG and KRS thank NYUAD's Center for Space Science. SMH acknowledges support from the Max-Planck partner group program and Ramanujan fellowship SB/S2/RJN-73. MFW acknowledges support from NASA grant NNX14AH84G.

\bibliographystyle{apj}
\bibliography{ms.bbl}

\begin{thebibliography}{}

\bibitem[\protect\citeauthoryear{{Christensen--Dalsgaard}}{{Christensen--Dalsgaard}}{2003}]{jcd_notes}
{Christensen--Dalsgaard}, J. 2003, Lecture Notes on Stellar Oscillations (Fifth
  ed.)

\bibitem[\protect\citeauthoryear{{Christensen-Dalsgaard}
  et~al.}{{Christensen-Dalsgaard} et~al.}{1996}]{jcd}
{Christensen-Dalsgaard}, J., et~al. 1996, Science, 272, 1286

\bibitem[\protect\citeauthoryear{{Dahlen}}{{Dahlen}}{1968}]{dahlen68}
{Dahlen}, F.~A. 1968, Geophysical Journal, 16, 329

\bibitem[\protect\citeauthoryear{{Dahlen} \& {Tromp}}{{Dahlen} \&
  {Tromp}}{1998}]{DT98}
{Dahlen}, F.~A.,  \& {Tromp}, J. 1998, {Theoretical Global Seismology}
  ({Princeton University Press})

\bibitem[\protect\citeauthoryear{{Duvall} et~al.}{{Duvall}
  et~al.}{1993}]{duvall}
{Duvall}, T.~L., Jr., {Jefferies}, S.~M., {Harvey}, J.~W.,  \& {Pomerantz},
  M.~A. 1993, \nat, 362, 430

\bibitem[\protect\citeauthoryear{{Gough}}{{Gough}}{1969}]{gough69}
{Gough}, D.~O. 1969, Journal of Atmospheric Sciences, 26, 448

\bibitem[\protect\citeauthoryear{{Gough}}{{Gough}}{1990}]{gough90}
{Gough}, D.~O. 1990, in Lecture Notes in Physics, Berlin Springer Verlag, Vol.
  367, Progress of Seismology of the Sun and Stars, ed. Y.~{Osaki} \&
  H.~{Shibahashi}, 283

\bibitem[\protect\citeauthoryear{{Greer} et~al.}{{Greer}
  et~al.}{2015}]{greer15}
{Greer}, B.~J., {Hindman}, B.~W., {Featherstone}, N.~A.,  \& {Toomre}, J. 2015,
  \apjl, 803, L17

\bibitem[\protect\citeauthoryear{{Hanasoge}, {Gizon}, \&
  {Sreenivasan}}{{Hanasoge} et~al.}{2016}]{arfm2016}
{Hanasoge}, S., {Gizon}, L.,  \& {Sreenivasan}, K.~R. 2016, Annual Review of
  Fluid Mechanics, 48, 191

\bibitem[\protect\citeauthoryear{{Hanasoge}, {Duvall}, \&
  {Sreenivasan}}{{Hanasoge} et~al.}{2012}]{hanasoge12_conv}
{Hanasoge}, S.~M., {Duvall}, T.~L., Jr.,  \& {Sreenivasan}, K.~R. 2012,
  {Proceedings of the National Academy of Sciences}, 109, 11928

\bibitem[\protect\citeauthoryear{{Hill}}{{Hill}}{1988}]{hill88}
{Hill}, F. 1988, \apj, 333, 996

\bibitem[\protect\citeauthoryear{{Larson} \& {Schou}}{{Larson} \&
  {Schou}}{2015}]{larson15}
{Larson}, T.~P.,  \& {Schou}, J. 2015, \solphys, 290, 3221

\bibitem[\protect\citeauthoryear{{Lavely} \& {Ritzwoller}}{{Lavely} \&
  {Ritzwoller}}{1992}]{lavely92}
{Lavely}, E.~M.,  \& {Ritzwoller}, M.~H. 1992, Philosophical Transactions of
  the Royal Society of London Series A, 339, 431

\bibitem[\protect\citeauthoryear{{Lynden-Bell} \& {Ostriker}}{{Lynden-Bell} \&
  {Ostriker}}{1967}]{ostriker67}
{Lynden-Bell}, D.,  \& {Ostriker}, J.~P. 1967, \mnras, 136, 293

\bibitem[\protect\citeauthoryear{{Miesch} et~al.}{{Miesch}
  et~al.}{2008}]{miesch_etal_08}
{Miesch}, M.~S., {Brun}, A.~S., {De Rosa}, M.~L.,  \& {Toomre}, J. 2008, \apj,
  673, 557

\bibitem[\protect\citeauthoryear{{Scherrer} et~al.}{{Scherrer}
  et~al.}{1995}]{scherrer95}
{Scherrer}, P.~H., et~al. 1995, \solphys, 162, 129

\bibitem[\protect\citeauthoryear{{Schou} et~al.}{{Schou} et~al.}{2012}]{hmi}
{Schou}, J., et~al. 2012, \solphys, 275, 229

\bibitem[\protect\citeauthoryear{Schulten \& Gordon}{Schulten \&
  Gordon}{1975}]{schulten1975}
Schulten, K.,  \& Gordon, R.~G. 1975, Journal of Mathematical Physics, 16, 1961

\bibitem[\protect\citeauthoryear{Vorontsov}{Vorontsov}{2011}]{vorontsov11}
Vorontsov, S.~V. 2011, Monthly Notices of the Royal Astronomical Society, 418,
  1146

\bibitem[\protect\citeauthoryear{{Woodard}}{{Woodard}}{2014}]{woodard14}
{Woodard}, M. 2014, \solphys, 289, 1085

\bibitem[\protect\citeauthoryear{{Woodard}}{{Woodard}}{2006}]{woodard06}
{Woodard}, M.~F. 2006, \apj, 649, 1140

\bibitem[\protect\citeauthoryear{{Woodard}}{{Woodard}}{2007}]{woodard07}
{Woodard}, M.~F. 2007, \apj, 668, 1189

\bibitem[\protect\citeauthoryear{{Woodard}}{{Woodard}}{2016}]{woodard16}
{Woodard}, M.~F. 2016, \mnras, 460, 3292

\bibitem[\protect\citeauthoryear{{Woodhouse}}{{Woodhouse}}{1980}]{woodhouse80}
{Woodhouse}, J.~H. 1980, Geophysical Journal, 61, 261

\end{thebibliography}
\appendix
\section{List of Symbols}\label{symbs}
\begin{eqnarray}
\gamma_\ell = \sqrt{\frac{2\ell+1}{4\pi}},\\
\Omega^\ell_N = \sqrt{\frac{1}{2}(\ell+N)(\ell-N+1)}.
\end{eqnarray}
We note that $\Omega^\ell_0 = \Omega^\ell_1$, and $\Omega^\ell_{-1} = \Omega^\ell_2.$

A useful identity involving Wigner-3$j$ symbols is
\begin{equation}
\int_0^{2\pi} d\phi \int_0^{\pi} d\theta\,\sin\theta (Y^{N'm'}_{\ell'})^* Y^{N{''}m{''}}_{\ell{''}} Y^{Nm}_\ell = 4\pi(-1)^{(N'-m')} \begin{pmatrix} \ell' & \ell{''} & \ell \\ -N' & N{''} & N\end{pmatrix} \begin{pmatrix}\ell' & \ell{''} & \ell \\ -m' & m{''} & m\end{pmatrix}.
\end{equation}
The Wigner-3$j$ symbol has a number of symmetry properties and the constituent parameters ($N, \ell$ etc.) must satisfy a variety of conditions for it to be non-zero. We do not describe these here but refer the reader to e.g. appendix~C, Section~d of \citet{lavely92} for a detailed list of properties.
The $B$-coefficient symbol that \citet{woodhouse80} and \citet{lavely92} significantly reduces the notational burden so we define and apply it in our analysis,
\begin{equation}
B^{(N)\pm}_{\ell' \ell{''}\ell } = \frac{(-1)^N}{2}(1\pm(-1)^{\ell+\ell'+\ell{''}})\left[\frac{(\ell'+N)!(\ell+N)!}{(\ell'-N)!(\ell-N)!}\right]^{1/2} \begin{pmatrix} \ell' & \ell{''} & \ell \\ -N & 0 & N\end{pmatrix}.
\end{equation}

\section{Useful Relations}
\begin{eqnarray}
Y^{Nm}_\ell(\theta,\phi) &=& D^\ell_{Nm}(\phi,\theta,0) = d^\ell_{Nm}(\theta) e^{im\phi},\nonumber\\
Y^m_\ell(\theta,\phi) &=& \gamma_\ell Y^{0m}_\ell(\theta,\phi),\nonumber\\
\frac{\partial Y^{Nm}_\ell}{\partial \theta} &=& \frac{1}{\sqrt2}(\Omega^\ell_N\,Y^{N-1,m}_\ell - \Omega^\ell_{N+1}\,Y^{N+1,m}_\ell),\nonumber\\
\frac{(N\cos\theta-m)}{\sin\theta}Y^{Nm}_\ell &=& \frac{1}{\sqrt2}(\Omega^\ell_{N}\,Y^{N-1,m}_\ell + \Omega^\ell_{N+1}\,Y^{N+1,m}_\ell).\label{recurse}
\end{eqnarray}
The following relations are particularly useful
\begin{eqnarray}
\frac{\partial Y^{0m}_\ell}{\partial \theta} &=& \frac{\Omega^\ell_0}{\sqrt2}(Y^{-1,m}_\ell - Y^{1,m}_\ell),\nonumber\\
-m\,\cosec\theta\,Y^{0m}_\ell &=& \frac{\Omega^\ell_0}{\sqrt2}(Y^{-1,m}_\ell + Y^{1,m}_\ell),\nonumber\\
\pth(Y^{-1,m}_\ell - Y^{1,m}_\ell) &=& \frac{1}{\sqrt2}\left[\Omega^\ell_{2}(Y^{-2m}_\ell + Y^{2m}_\ell) -2\Omega^\ell_0 Y^{0m}_\ell\right]\nonumber\\
\pth(Y^{-1,m}_\ell + Y^{1,m}_\ell) &=&  \frac{\Omega^\ell_{2}}{\sqrt2}\left(Y^{-2m}_\ell - Y^{2m}_\ell \right),\nonumber\\
m\,\cosec\theta\,(Y^{-1m}_\ell - Y^{1m}_\ell) +{\cot\theta}(Y^{-1m}_\ell + Y^{1m}_\ell) &=& \frac{\cos\theta -m}{\sin\theta} Y^{1m}_\ell -\frac{-\cos\theta -m}{\sin\theta} Y^{-1m}_\ell ,\nonumber\\
&=& \frac{\Omega^\ell_{2}}{\sqrt2}\left(Y^{2m}_\ell - Y^{-2m}_\ell \right),\nonumber\\
\cot\theta(Y^{-1,m}_\ell - Y^{1,m}_\ell) +m\,\cosec\theta\,(Y^{-1,m}_\ell + Y^{1,m}_\ell) &=& -\frac{-\cos\theta -m}{\sin\theta} Y^{-1m}_\ell - \frac{\cos\theta -m}{\sin\theta} Y^{1m}_\ell\nonumber\\
&=& -\frac{1}{\sqrt2}\left[ \Omega^\ell_{2}(Y^{2m}_\ell + Y^{-2m}_\ell) + 2\Omega^\ell_0 Y^{0m}_\ell\right].\nonumber\\
\label{specuse}
\end{eqnarray}

We note the following relationships,
\begin{eqnarray}
\frac{\partial\rhat}{\partial r} &=& 0,\,\,\,\, \frac{\partial\rhat}{\partial\theta}=\that,\,\,\,\, \frac{\partial\rhat}{\partial\phi}=\phat\,\sin\theta,\nonumber\\
\frac{\partial\that}{\partial r} &=& 0,\,\,\,\, \frac{\partial\that}{\partial\theta}=-\rhat,\,\,\,\, \frac{\partial\that}{\partial\phi}=\phat\,\cos\theta,\nonumber\\
\frac{\partial\phat}{\partial r} &=& 0,\,\,\,\, \frac{\partial\phat}{\partial\theta}= 0,\,\,\,\, \frac{\partial\phat}{\partial\phi}=-\rhat\,\sin\theta -\that\,\cos\theta.\label{coordinate}
\end{eqnarray}

This allows us to write tensor $\bT = \bnabla\bxi_k$ so
\begin{eqnarray}
T_{rr} &=&  \gamma_\ell\,\dU(r)\,Y^{0m}_\ell,\nonumber\\
T_{r\theta} &=&   \frac{\gamma_\ell}{\sqrt2}\,\dV(r)\,\Omega^\ell_0\,(Y^{-1,m}_\ell - Y^{1,m}_\ell),\nonumber\\
T_{r\phi} &=& -i\frac{\gamma_\ell}{\sqrt2}\,\dV(r)\,\Omega^\ell_0\,(Y^{-1,m}_\ell + Y^{1,m}_\ell),\nonumber\\
T_{\theta r} &=& r^{-1}\gamma_\ell\,\left[U(r) \pth Y^{0m}_\ell - \frac{V(r)}{\sqrt2}\,\Omega^\ell_0\,(Y^{-1,m}_\ell - Y^{1,m}_\ell)\right],\nonumber\\
T_{\theta \theta} &=& r^{-1}{\gamma_\ell}\left[U(r) Y^{0m}_\ell +\frac{V(r)}{\sqrt2}\,\Omega^\ell_0\,\pth(Y^{-1,m}_\ell - Y^{1,m}_\ell)\right],\nonumber\\
T_{\theta \phi} &=& -ir^{-1}\frac{\gamma_\ell}{\sqrt2}\,V(r)\,\Omega^\ell_0\,\pth(Y^{-1,m}_\ell + Y^{1,m}_\ell),\nonumber\\
T_{\phi r} &=& ir^{-1}\gamma_\ell \left[m\,U(r)\,\cosec\theta\,Y^{0m}_\ell +\frac{V(r)}{\sqrt2}\,\Omega^\ell_0\,(Y^{-1m}_\ell + Y^{1m}_\ell)\right],\nonumber\\
T_{\phi \theta} &=& ir^{-1}\frac{\gamma_\ell}{\sqrt2}\,V(r)\,\Omega^\ell_0 \left[m\,\cosec\theta\,(Y^{-1,m}_\ell - Y^{1,m}_\ell) +{\cot\theta}(Y^{-1m}_\ell + Y^{1m}_\ell)\right],\nonumber\\
T_{\phi\phi} &=& r^{-1}\gamma_\ell\left[U(r) Y^{0m}_\ell + \Omega^\ell_0\frac{V(r)}{\sqrt2}[\cot\theta(Y^{-1,m}_\ell - Y^{1,m}_\ell) +m\,\cosec\theta\,(Y^{-1,m}_\ell + Y^{1,m}_\ell)] \right].\nonumber\\\label{Tsimple}
\end{eqnarray}

\section{Properties of the Adjoints}\label{adjoint.props}
Consider the flow operator $\delta\op = -2i\omega\bu\cdot\bnabla$, and keeping in mind that because $\bu$ is real in the spatio-temporal domain, $\bu^*(\omega) = \bu(-\omega)$. Anelasticity \citep{gough69} allows us to write $\bnabla\cdot(\rho\bu) = 0$. Recalling the definition from equation~(\ref{coupledef}), we have for $\Lambda^{k'}_k(\omega)$,
\begin{eqnarray}
\Lambda^{k'}_k(\omega) = 2\int_\odot d\br\,i\omega\rho\,\bs^*_{k'}\cdot\bu\cdot\bnabla\bs_k = 
\int_\odot \,[ 2\bnabla\cdot(i\omega\rho\bu\,\bs^*_{k'}\cdot\bs_{k}) - 2i\omega\rho\,\bs_k\cdot\bu\cdot\bnabla\bs^*_{k'}]\,d\br,\nonumber\\
= \left[\int_\odot\, 2i\omega\rho\,\bs^*_k\cdot(\bu^*\cdot\bnabla)\bs_{k'}\,d\br\right]^* = \Lambda^{k*}_{k'}(-\omega).
\end{eqnarray}
Similarly, for the scattering due to thermal anomalies, where $\delta\op = -\bnabla[\delta(\rho c^2)\,\bnabla\cdot ]$, and keeping in mind that $c (\br,-\omega) = c^* (\br,\omega)$, we have
\begin{eqnarray}
&&\Lambda^{k'}_k(\omega) = \int_\odot d\br\,\bs^*_{k'}\cdot\bnabla[\delta(\rho c^2)\bnabla\cdot\bs_k] = 
\int_\odot\, \{\bnabla\cdot[\bs^*_{k'}\,\delta(\rho c^2)\bnabla\cdot\bs_k] - \delta(\rho c^2)\bnabla\cdot\bs^*_{k'}\,\bnabla\cdot\bs_k\}\,d\br\nonumber\\
&&= \int_\odot d\br\, \bs_k\cdot\bnabla[\delta(\rho c^2)\bnabla\cdot\bs^*_{k'}] = \left\{\int_\odot\, \bs^*_k\cdot\bnabla[\delta(\rho c^2)^*\bnabla\cdot\bs_{k'}]\,d\br \right\}^*= \Lambda^{k*}_{k'}(-\omega).
\end{eqnarray}

\section{Fourier convention}\label{fourier}
The following forward and inverse transforms connect the function $f(t)$ and its Fourier transform $\hat{f}(\omega)$,
\begin{equation}
\hat{f}(\omega) = \frac{1}{2\pi}\int dt\, f(t)\, e^{i\omega t},
\end{equation}
and
\begin{equation}
f(t) = \int d\omega\, \hat{f}(\omega)\, e^{-i\omega t}.
\end{equation}
The $\delta$ function is defined thus,
\begin{equation}
\int dt\,e^{i\omega t} = 2\pi\,\delta(\omega).
\end{equation}
A product in time domain $h(t) = f(t) g(t)$ transforms to the following convolution in Fourier domain,
\begin{equation}
\hat{h}(\omega) = \int d\omega'\,\hat{f}(\omega')\,\hat{g}(\omega-\omega').\label{convint}
\end{equation}

\end{document}